\documentclass[preprint,12pt]{elsarticle}

\usepackage{amsmath,amssymb,amsfonts}
\usepackage[table]{xcolor}
\usepackage{amssymb}
\usepackage{graphicx}
\usepackage{wrapfig}
\usepackage{subfig}
\usepackage{url}
\usepackage{verbatim}
\usepackage{multirow}
\usepackage{paralist}
\usepackage{mathptmx}
\usepackage{times}
\usepackage{dirtree}
\usepackage{tabularx}
\usepackage{tikz}
\usepackage[T1]{fontenc}
\usepackage{pgfplots}
\usepackage{todonotes}
\usepackage[pdfsubject={Robotic Process Automation, Process Mining},
                            pdftitle={},
                            pdfauthor={},
                            pdfkeywords={},
                            colorlinks=false,
                            pdfborder={0.25 0.25 0.25}]{hyperref}

\usepackage[capitalise,nameinlink]{cleveref}
\usepackage[ruled,vlined,boxed,linesnumbered]{algorithm2e}
\SetKwRepeat{Do}{do}{while}
\SetKwInOut{Input}{input}
\SetKwInOut{Output}{output}
\SetKw{Break}{break}

\journal{Information Systems}

\begin{document}

\sloppy

\begin{frontmatter}

\title{Discovering Executable Routine Specifications From User Interaction Logs}

\author[1,2]{Volodymyr Leno}
\ead{leno@ut.ee}
\author[2]{Adriano Augusto}
\ead{a.augusto@unimelb.edu.au}
\author[1]{Marlon Dumas}
\ead{marlon.dumas@ut.ee}
\author[2]{Marcello La Rosa}
\ead{marcello.larosa@unimelb.edu.au}
\author[3]{Fabrizio Maria Maggi}
\ead{maggi@inf.unibz.it}
\author[2]{Artem Polyvyanyy}
\ead{artem.polyvyanyy@unimelb.edu.au}

\address[1]{University of Tartu, Liivi 2, 50409, Tartu, Estonia}
\address[2]{University of Melbourne, Parkville, VIC, 3010, Australia}
\address[3]{Free University of Bozen-Bolzano, Piazza Universita, 1, 39100, Italy}

\begin{abstract}
Robotic Process Automation (RPA) is a technology to automate routine work such as copying data across applications or filling in document templates using data from multiple applications. 
RPA tools allow organizations to automate a wide range of routines. 
However, identifying and scoping routines that can be automated using RPA tools is time consuming. Manual identification of candidate routines via interviews, walk-throughs, or job shadowing allow analysts to identify the most visible routines, but these methods are not suitable when it comes to identifying the long tail of routines in an organization.
This article proposes an approach to discover automatable routines from logs of user interactions with IT systems and to synthetize executable specifications for such routines. 
The approach starts by discovering frequent routines at a control-flow level (candidate routines). It then determines which of these candidate routines are automatable and it synthetizes an executable specification for each such routine. Finally, it identifies semantically equivalent routines so as to produce a set of non-redundant automatable routines. The article reports on an evaluation of the approach using a combination of synthetic and real-life logs. The evaluation results show that the approach can discover automatable routines that are known to be present in a UI log, and that it identifies automatable routines that users recognize as such in real-life logs.
\end{abstract}





\begin{keyword}
Robotic Process Automation \sep Robotic Process Mining \sep UI log
\end{keyword}

\end{frontmatter}

\newtheorem{definition}{Definition}

\newcommand{\sbt}{\,\begin{picture}(-1,1)(-1,-3)\circle{3}\end{picture}\ \ }

\newcommand{\nui}{\bar{u}}
\newcommand{\nuilog}{\bar{\Sigma}}

\newcommand{\reimbursement}{{\sc RT}}
\newcommand{\studentRecord}{{\sc SR}}
\newcommand{\scholarshipA}{{\sc S1}}
\newcommand{\scholarshipB}{{\sc S2}}

\newcommand{\srx}{{\sc SRRT\textsubscript{$+$}}}
\newcommand{\sry}{{\sc RTSR\textsubscript{$+$}}}
\newcommand{\srz}{{\sc SRRT\textsubscript{$\parallel$}}}
\newcommand{\srk}{{\sc RTSR\textsubscript{$\parallel$}}}

\newcommand{\crset}{\mathcal{C}_{\Sigma}}
\newcommand{\riset}{\mathcal{R}_{c_{i}}}
\newcommand{\rset}{\mathcal{R}}


\section{Introduction}
\label{sec:intro}

Robotic Process Automation (RPA) allows organizations to improve their processes by automating repetitive sequences of interactions between a user and one or more software applications (a.k.a.\ routines).
Using this technology, it is possible to automate data entry, data transfer, and verification tasks, particularly when such tasks involve multiple applications.
To exploit this technology, organizations need to identify routines that are amenable to automation~\cite{leopold2018identifying}. 
This can be achieved via interviews, walk-throughs, job shadowing, or by examining documented procedures~\cite{leopold2018identifying}. These approaches are not always cost-efficient in large organizations, as routines tend to be scattered across the process landscape. 

To tackle this gap, several research studies have proposed techniques to analyze User Interaction (UI) logs in order to discover repetitive routines that are amenable to automation via RPA~\cite{jimenez2019method,bosco2019,gao2019automated,leno2020aaai,DBLP:conf/bpm/AgostinelliLMM20}. However, existing approaches in this space make various assumptions that limit their applicability. 

First, all of the existing approaches for discovering frequent and/or automatable routines from UI logs assume that the UI log consists of a set of traces (segments) of a task that is presupposed to contain one or more routines. 
In practice, however, UI logs are not segmented. 
Instead, a recording of a working session consists of a single sequence of actions encompassing many instances of one or more routines, interspersed with other events that may not be part of any routine. 

Second, most of the existing approaches ~\cite{jimenez2019method,bosco2019,gao2019automated} discover frequent routines and/or automatable routines, but they do not produce an executable routine specification. 

Third, existing approaches do not take into account the fact that the same routine may be performed differently (albeit equivalently) by different workers, or sometimes even by the same worker. In other words, existing approaches may produce redudant routines as output.


This article addresses these gaps by presenting an approach to discover automatable routines from unsegmented UI logs. The approach splits the unsegmented UI log into a set of segments, each representing a sequence of steps that appears frequently in the unsegmented UI log. 
It then applies sequential pattern mining techniques to find candidate routines for automation and evaluates their automatability.  
For each automatable routine, the approach synthesizes an executable routine specification, 
which can be compiled into an RPA bot.  This bot can then be executed by an RPA tool to replicate the underlying routine automatically. 

The proposed approach has been implemented as an open-source prototype called Robidium~\cite{LenoDPRDM20}. 
Using this implementation, we have evaluated the proposed approach on synthetic and real-life UI logs in terms of its execution times and its ability to accurately discover routines from an UI log.


This article is an extended and revised version of a conference paper~\cite{DBLP:conf/icpm/LenoADRMP20}. 
The conference version focused on the discovery of frequently repeated routines from unsegmented UI logs (i.e.\ candidate routines). 
This article extends this initial approach in two ways. First, this article presents an approach to post-process the identified candidate routines in order to assess their automatability and, in case a routine is fully automatable, to generate an executable routine specification. Second, this article proposes a method to identify semantically equivalent routines, so as to produce a non-redundant set of automable routines.

This article provides a concrete realization of a high-level architecture for discovering automatable routines from UI logs, sketched in~\cite{lenobise20}. To this end, the article proposes concrete techniques to implement each of the building blocks in~\cite{lenobise20}, except for the UI log recording step, which is documented in~\cite{DBLP:conf/bpm/LenoPRDM19}.


The article is structured as follows.  Section \ref{sec:related} provides an overview of related work.  Section \ref{sec:approach} describes the approach, while Section \ref{sec:evaluation} reports the results of the evaluation. Finally, Section \ref{sec:conclusion} concludes the paper and discusses the directions for future work. 

\section{Related work}
\label{sec:related}

The problem addressed by this article is denominated as Robotic Process Mining (RPM) in~\cite{lenobise20}.  RPM is a family of methods to discover repetitive routines performed by employees during their daily work,  and to turn such routines into software scripts that emulate their execution.  The first step in an RPM pipeline is to record the interactions between one or more workers and one or more software applications~\cite{DBLP:conf/bpm/LenoPRDM19}.  The recorded data is represented as a UI log -- a sequence of user interactions (herein called UIs), such as selecting a cell in a spreadsheet or editing a text field in a form.  The UI log may be filtered to remove irrelevant UIs (e.g., misclicks).  Next, it may be decomposed into segments (segmentation).  The discovered segments are then scanned to identify routines that occur frequently across these segments.  Finally, the resulting frequent routines (a.k.a.\ candidate routines) are analyzed in order to identify those that are automatable and to derive executable routine specifications.

In this section, we review previous research related to the three core research challenges of RPM identified in~\cite{lenobise20}: UI log segmentation, discovery of frequent (candidate) routines and discovery of automatable routines.

\subsection{UI Log Segmentation}

Given a UI log (i.e., a sequence of UIs), segmentation consists in identifying non-overlapping subsequences of UIs, 
namely \emph{segments}, such that each subsequence represents the execution of a task performed by an employee from start to end. 
In other words, segmentation searches for repetitive patterns in the UI log. 
In an ideal scenario, we would observe only one unique pattern (the task execution) repeated a finite number of times. 
However, in reality, this scenario is unlikely to materialize. 
Instead, it is reasonable to assume that an employee performing X-times the same task would make some mistakes or introduce variance in how the task is performed.

The problem of segmentation is similar to periodic pattern mining on time series. 
While several studies addressed the latter problem over the past decades~\cite{cao2007discovery,zhu2017matrix}, 
most of them require information regarding the length of the pattern to discover or assume a natural period to be available (e.g., hour, day, week). 
This makes the adaptation of such techniques to solve the problem of segmentation challenging unless periodicity and pattern length are known a priori.

Under the same class of problems, we find web session reconstruction~\cite{spiliopoulou2003framework}, 
whose goal is to identify the beginning and the end of web navigation sessions in server log data (e.g., streams of clicks and web page navigation)~\cite{spiliopoulou2003framework}. 
Methods for session reconstruction are usually based on heuristics that rely on structural organization of web sites or time intervals between events. 
The former approach covers only the cases when all the user interactions are performed in the web applications,
while the latter approach assumes that users make breaks in-between two consecutive segments -- in our case, two routine instances.

Lastly, segmentation also relates to the problem of correlation of event logs for process mining. 
In such logs, each event should normally include an identifier of a process instance (case identifier), a timestamp, an activity label, and possibly other attributes. 
When the events in an event log do not contain explicit case identifiers, they are said to be uncorrelated.  
Various methods have been proposed to extract correlated event logs from uncorrelated ones. 
However, existing methods in this field either assume that a process model is given as input~\cite{DBLP:conf/caise/BayomieAE16} or that the underlying process is acyclic~\cite{DBLP:conf/bpm/FerreiraG09}. 
Both of these assumptions are unrealistic in our setting: a process model is not available since we are precisely trying to identify the routines in the log, and a routine may contain repetition. 

Recent work on UI log segmentation~\cite{DBLP:conf/icpm/Agostinelli20} proposes to use trace alignment between the logs and the corresponding interaction models to identify the segments. In practice, however, such interaction models are not available beforehand. In this article, we outline a segmentation approach that does not require any models as inputs nor does it require that the user specifies one or more explicit delimiters  between segments (e.g.\ that the user specifies that a given symbol X represents the start and/or the end of a segment).


\subsection{Frequent Routine Discovery}


Dev and Liu~\cite{DBLP:conf/iui/DevL17} have noted that the problem of routine identification from (segmented) UI logs can be mapped to that of frequent pattern mining,  a well-known problem in the field of data mining~\cite{han2007frequent}. 
Indeed, the goal of routine identification is to identify repetitive (frequent) sequences of interactions, which can be represented as symbols. 
In the literature, several algorithms are available to mine frequent patterns from sequences of symbols.
Depending on their output, we can distinguish two types of frequent pattern mining algorithms: 
those that discover only exact patterns~\cite{lee2004efficient,ohlebusch2015alphabet} (hence vulnerable to noise), 
and those that allow frequent patterns to have gaps within the sequence of symbols~\cite{wang2004bide,fumarola2016clofast} (hence noise-resilient).

Depending on their input, we can distinguish between algorithms that operate on a collection of sequences of symbols and those that discover frequent patterns from a single long sequence of symbols~\cite{ohlebusch2015alphabet}. 
The former algorithms can be applied to segmented UI logs, while the latter can be applied directly to unsegmented ones. 
However, techniques that identify patterns from a single sequence of symbols only scale up when identifying exact patterns.
While such approaches discover the frequently repeated routines, they do not analyze whether they are automatable. 
In other words, these approaches focus on the discovery of the control-flow models instead of executable specifications. 

The identification of frequent routines from sequences of actions is related to the problem of Automated Process Discovery (APD) \cite{DBLP:journals/tkde/AugustoCDRMMMS19}, 
which has been studied in the field of process mining. 
Recent works~\cite{DBLP:conf/bpm/Geyer-Klingeberg18,jimenez2019method} show that RPA can benefit from process mining. 
In particular, the work in \cite{jimenez2019method} proposes to apply traditional APD techniques to discover process models of routines captured in UI logs. 
However, traditional APD techniques discover control-flow models, while, in the context of RPA, 
we seek to discover executable specifications that capture the mapping between the outputs and the inputs of the actions performed during a routine.


\subsection{Discovery of Automatable Routines}

The discovery of automatable sequences of user interactions has been widely studied in the context of Web form and table auto-completion. For example, Excel's Flash Fill feature detects string patterns in the values of the cells in a spreadsheet and uses these patterns for auto-completion~\cite{DBLP:conf/popl/Gulwani11}. However, auto-completion techniques focus on identifying repetitions of keystrokes (sequences of characters). In this article, we look at routines that involve transfering data across fields in one or more applications as well as editing field values.

The discovery of data transfer routines that are amenable for RPA automation has been addressed in~\cite{bosco2019}. This latter paper proposes a technique to discover sequences of actions such that the inputs of each action in the sequence (except the first one) can be derived from the data observed in previous actions. However, this technique can only discover perfectly sequential routines, 
and is hence not resilient to variability in the order of the actions, 
whereas in reality, different users may perform the actions in a routine in a different order.

Another technique for routine identification~\cite{leopold2018identifying} attempts to identify candidate routines from textual documents -- an approach that is suitable for earlier stages of routine identification and could be used to determine which processes or tasks could be recorded and analyzed in order to identify routines.

In \cite{DBLP:conf/bpm/AgostinelliLMM20} the authors present an approach to automatically discover routines from UI logs and automate them in the form of scripts. 
This approach, however, assumes that all the actions within a routine are automatable.
In practice, it is possible that some actions have to be performed manually, and they can not be automated. 

The approach presented in \cite{gao2019automated} aims at extracting rules from segmented UI logs that can be used to fill in forms automatically.
However, this approach only discovers branching conditions that specify whether a certain activity has to be performed or not (e.g., check the box of the form). 
It focuses only on the copy-paste operations and does not identify more complex manipulations. 

In previous work~\cite{leno2020aaai}, we mapped the problem of discovering routines related to the data transferring to the problem of discovering data transformations. In this paper, we reuse this idea and extend it to tackle the problem of assessing if and to what extent a frequent (candidate) routine is automatable, and if such, producing an executable specification.


\section{Approach}
\label{sec:approach}

\begin{figure*}[htb]
\centering
\includegraphics[scale=0.6]{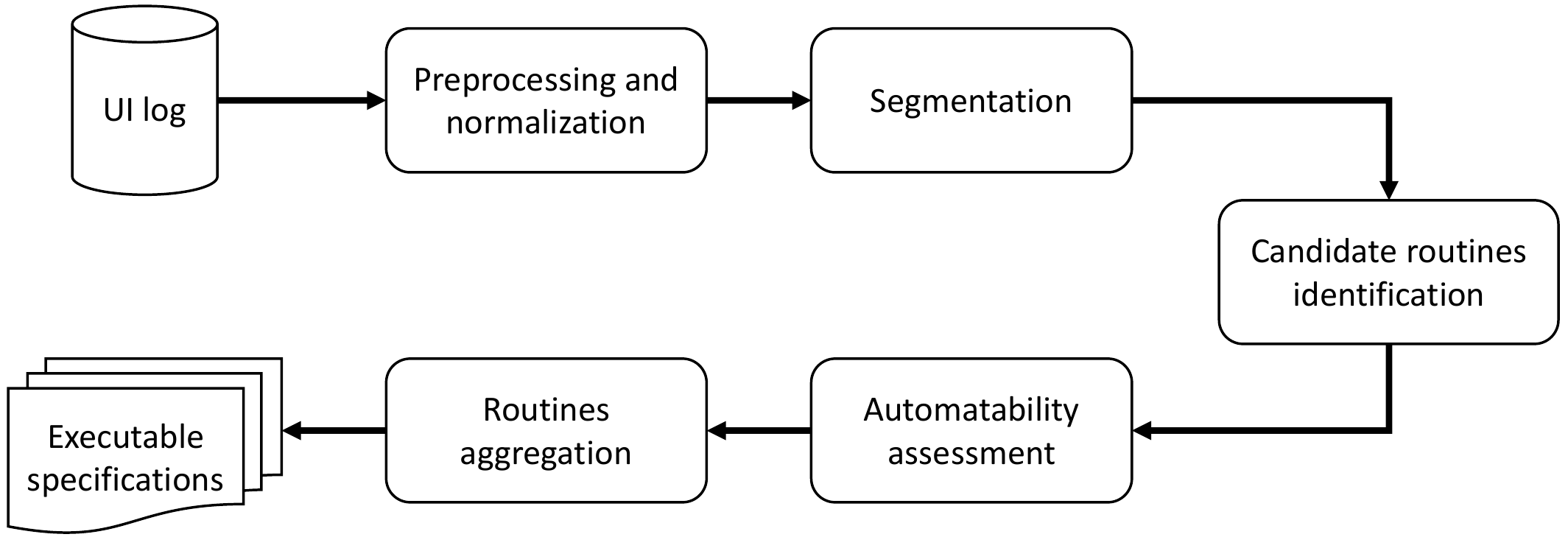}
\caption{Outline of the proposed approach}
\label{fig:approach}
\end{figure*}

In this section, we describe our approach for discovering executable routine specifications from User Interaction (UI) logs.
We adhere to the RPM pipeline proposed by Leno et al.~\cite{lenobise20}, which we implemented in five macro steps (see Figure~\ref{fig:approach}):
i) \emph{preprocessing and normalization}; ii) \emph{segmentation}; iii) \emph{candidate routine identification}; iv) \emph{automatability assessment}; v) \emph{routines aggregation}.

Our approach takes as input a UI log, which is a chronologically ordered sequence of UIs between a worker and computer-based applications.
In this paper, we assume that the applications used by the worker are either spreadsheet management applications or web browser applications. 
A UI log is usually recorded during the execution of the worker's daily tasks using specialized logging tools,
for example, the \emph{Action Logger} tool~\cite{DBLP:conf/bpm/LenoPRDM19}.

An example of a UI log is provided in Table~\ref{tab:uiLog}.
Each row of Table~\ref{tab:uiLog} captures one UI (e.g., clicking a button or copying the content of a cell).
Each UI is characterized by a \emph{timestamp}, a \emph{type}, and a set of \emph{parameters}, or \emph{payload} (e.g., application, button's label and value of a field).
The payload of a UI is not standardized, and depends on the UI type and application.
Consequently, the UIs recorded in the same log may have different payloads.
For example, the payload of UIs performed within a spreadsheet contains information regarding the spreadsheet name and the location of the target cell (e.g., cell row and column).
In contrast, the payload of the UIs performed in a web browser contains information regarding the webpage URL, the name and identifier of the UI's target HTML element and its value (if any); -- see Table~\ref{tab:uiLog} rows 1 and 2.

\begingroup
\begin{table}[tbp]
\makebox[0.98\textwidth][c]{
\centering
{\tiny{
    \begin{tabular}{c|c|c|c|c|c|c|c|c}
    & \textbf{UI}
    & \textbf{UI}
    & \multicolumn{6}{|c}{\textbf{Payload}}
    \\\cline{4-9}
	
	\textbf{Row}
    & \textbf{Timestamp}
    & \textbf{Type}
    & $\boldmath{P_1}$
    & $\boldmath{P_2}$
    & $\boldmath{P_3}$
    & $\boldmath{P_4}$
    & $\boldmath{P_5}$
    & $\boldmath{P_6}$\\\hline

    1 & 2019-03-03T19:02:18 & Click button (Web) & https://unimelb.edu.au & New Record & newRecord & button & -- & --  \\\hline
    2 & 2019-03-03T19:02:20 & Select cell (Excel) & StudentRecords & Sheet1 & A & 2 & Albert Rauf & -- \\\hline 
    3 & 2019-03-03T19:02:23 & Copy cell (Excel) & StudentRecords & Sheet1 & A & 2 & Albert Rauf & Albert Rauf \\\hline 
    4 & 2019-03-03T19:02:25 & Select field (Web) & https://unimelb.edu.au & Full Name & name & ``'' & -- & -- \\\hline 
    5 & 2019-03-03T19:02:26 & Paste (Web) & https://unimelb.edu.au & Full Name & name & ``'' & Albert Rauf & -- \\\hline 
    6 & 2019-03-03T19:02:28 & Edit field (Web) & https://unimelb.edu.au & Full Name & name & text & Albert Rauf & -- \\\hline
    7 & 2019-03-03T19:02:30 & Select cell (Excel) & StudentRecords & Sheet1 & B & 2 & 11/04/1986 & -- \\\hline 
    8 & 2019-03-03T19:02:31 & Copy cell (Excel) & StudentRecords & Sheet1 & B & 2 & 11/04/1986 & 11/04/1986 \\\hline  
    9 & 2019-03-03T19:02:34 & Select field (Web) & https://unimelb.edu.au & Date & date & ``'' & -- & -- \\\hline 
   10 & 2019-03-03T19:02:37 & Paste (Web) & https://unimelb.edu.au & Date & date & ``'' & 11/04/1986 & -- \\\hline 
   11 & 2019-03-03T19:02:40 & Edit field (Web) &  https://unimelb.edu.au & Date & date & text & 11-04-1986 & --  \\\hline
   12 & 2019-03-03T19:07:30 & Select cell (Excel) & StudentRecords & Sheet1 & C & 2 & +61 043 512 4834 & -- \\\hline 
   13 & 2019-03-03T19:07:33 & Copy cell (Excel) & StudentRecords & Sheet1 & C & 2 & +61 043 512 4834 & +61 043 512 4834 \\\hline 
   14 & 2019-03-03T19:07:40 & Select field (Web) & https://unimelb.edu.au & Phone & phone & ``'' & -- & -- \\\hline 
   15 & 2019-03-03T19:07:46 & Paste (Web) & https://unimelb.edu.au & Phone & phone & ``'' & +61 043 512 4834 & -- \\\hline 
   16 & 2019-03-03T19:07:48 & Edit field (Web) & https://unimelb.edu.au & Phone & phone & text & 043-512-4834 & -- \\\hline
   17 & 2019-03-03T19:07:50 & Select cell (Excel) & StudentRecords & Sheet1 & D & 2 & Germany & --\\\hline 
   18 & 2019-03-03T19:07:52 & Copy cell (Excel) & StudentRecords & Sheet1 & D & 2 & Germany & Germany \\\hline 
   19 & 2019-03-03T19:07:55 & Select field (Web) & https://unimelb.edu.au & Country of residence & country & ``'' & -- & -- \\\hline 
   20 & 2019-03-03T19:07:57 & Paste (Web) & https://unimelb.edu.au & Country of residence & country & ``'' & Germany & -- \\\hline 
   21 & 2019-03-03T19:07:59 & Edit field (Web) & https://unimelb.edu.au & Country of residence & country & text & Germany & --  \\\hline
   22 & 2019-03-03T19:08:02 & Edit field (Web) & https://unimelb.edu.au & Student status & status & select & Domestic & -- \\\hline   
   23 & 2019-03-03T19:08:05 & Edit field (Web) & https://unimelb.edu.au & Student status & status & select & International & -- \\\hline   
   24 & 2019-03-03T19:08:08 & Click button (Web) & https://unimelb.edu.au & Submit & submit & submit & -- & --\\\hline
   25 & 2019-03-03T19:08:12 & Click button (Web) & https://unimelb.edu.au & New Record & newRecord & button & -- & -- \\\hline
   26 & 2019-03-03T19:08:15 & Select cell (Excel) & StudentRecords & Sheet1 & B & 3 & 20/06/1987 & -- \\\hline  
   27 & 2019-03-03T19:08:18 & Copy cell (Excel) & StudentRecords & Sheet1 & B & 3 & 20/06/1987 & 20/06/1987 \\\hline  
   28 & 2019-03-03T19:08:21 & Select field (Web) & https://unimelb.edu.au & Date & date & ``'' & -- & -- \\\hline 
   29 & 2019-03-03T19:08:26 & Paste (Web) & https://unimelb.edu.au & Date & date & ``'' & 20/06/1987 & -- \\\hline 
   30 & 2019-03-03T19:08:28 & Edit field (Web) & https://unimelb.edu.au & Date & date & text & 20-06-1987 & -- \\\hline
   31 & 2019-03-03T19:08:32 & Select cell (Excel) & StudentRecords &  Sheet1 & C & 3 & +61 519 790 1066 & -- \\\hline 
   32 & 2019-03-03T19:08:34 & Copy cell (Excel) & StudentRecords &  Sheet1 & C & 3 & +61 519 790 1066 & +61 519 790 1066 \\\hline 
   33 & 2019-03-03T19:08:36 & Select field (Web) & https://unimelb.edu.au & Phone & phone & ``'' & -- & -- \\\hline 
   34 & 2019-03-03T19:08:38 & Paste (Web) & https://unimelb.edu.au & Phone & phone & ``'' & +61 519 790 1066 & -- \\\hline 
   35 & 2019-03-03T19:08:39 & Edit field (Web) & https://unimelb.edu.au & Phone & phone & text & 519-790-1066 & -- \\\hline
   36 & 2019-03-03T19:08:40 & Select cell (Excel) & StudentRecords & Sheet1 & A & 3 & Audrey Backer & -- \\\hline 
   37 & 2019-03-03T19:08:41 & Copy cell (Excel) & StudentRecords & Sheet1 & A & 3 & Audrey Backer & Audrey Backer \\\hline 
   38 & 2019-03-03T19:08:42 & Select field (Web) & https://unimelb.edu.au & Full Name & name & ``'' & -- & -- \\\hline 
   39 & 2019-03-03T19:08:44 & Paste (Web) & https://unimelb.edu.au & Full Name & name & ``'' & Audrey Backer & -- \\\hline 
   40 & 2019-03-03T19:08:46 & Edit field (Web) & https://unimelb.edu.au & Full Name & name & text & Audrey Backer & -- \\\hline
   41 & 2019-03-03T19:08:50 & Select cell (Excel) & StudentRecords & Sheet1 & D & 2 & Germany & -- \\\hline 
   42 & 2019-03-03T19:08:52 & Copy cell (Excel) & StudentRecords & Sheet1 & D & 2 & Germany & Germany \\\hline 
   43 & 2019-03-03T19:08:58 & Select cell (Excel) & StudentRecords & Sheet1 & D & 3 & Australia & -- \\\hline 
   44 & 2019-03-03T19:09:01 & Copy cell (Excel) & StudentRecords & Sheet1 & D & 3 & Australia & Australia \\\hline 
   45 & 2019-03-03T19:09:05 & Select field (Web) & https://unimelb.edu.au & Country of residence & country & ``'' & -- & -- \\\hline 
   46 & 2019-03-03T19:09:08 & Paste (Web) & https://unimelb.edu.au & Country of residence & country & ``'' & Australia & -- \\\hline 
   47 & 2019-03-03T19:09:10 & Edit field (Web) & https://unimelb.edu.au & Country of residence & country & text & Australia & -- \\\hline
   48 & 2019-03-03T19:09:14 & Edit field (Web) & https://unimelb.edu.au & Student status & status & select & Domestic & -- \\\hline   
   49 & 2019-03-03T19:09:20 & Click button (Web) & https://unimelb.edu.au & Submit & submit & submit & -- & -- \\\hline
  \dots & \dots & \dots & \dots & \dots & \dots & \dots & \dots & \dots \\\hline
 \end{tabular}
	}}}
   \caption{Fragment of a user interaction log}\label{tab:uiLog}
   \vspace{-2mm}
\end{table}
\endgroup

\newpage

Our approach analyzes the log to identify and output a collection of \emph{executable routine specifications}.
Each routine specification is a pair ($c$, $\Lambda$),
where $c$ is a sequence of UIs, or a \emph{candidate routine}, and $\Lambda$ is a set of \emph{data transformation steps}.
Each \emph{data transformation step} is a triplet that specifies: i) variables from which the data was read, ii) variables to which the data was written,
and iii) a function capturing the data transformation (if any occurs).
Such routine specifications can be compiled into software bots that can be deployed on a tool like UiPath,~\footnote{A commercial tool available at www.uipath.com} which would be able to automatically replicate the routine.

In the following, we describe step-by-step how we generate a collection of executable routine specifications from an input UI log.


\subsection{Preprocessing and Normalization}
\label{sec:preprocessing}

Before diving into the details of this step, we formally define the concepts of a \emph{user interaction} and \emph{user interaction log},
which we will refer to throughout this and the following sections.

\begin{definition}[\textbf{User interaction (UI)}]
A \emph{user interaction (UI)} is a tuple $u = (t, \tau, P_{\tau}, Z, \phi)$, where:
$t$ is a timestamp;
$\tau$ is a UI type;
$P_{\tau}$ is a set of parameters, or \emph{payload};
$Z$ is a set of parameter values; and
$\phi : P_{\tau} \rightarrow Z$ is a value assignment function.
\end{definition}

Table~\ref{tab:uiParam} shows UIs and their associated payloads recorded by the Action Logger tool~\cite{DBLP:conf/bpm/LenoPRDM19}.
The UIs are logically grouped, based on their type, into three groups:
\emph{navigation}; \emph{read}; and \emph{write} UIs.
We assume that every UI is an \emph{instantiation} of one of the UI types from Table~\ref{tab:uiParam},
with every parameter assigned with a specific value.

\begin{definition}[\textbf{User interaction log}]
A user interaction log $\Sigma$ is a sequence of UIs $\Sigma = \langle u_1, u_2, \dots, u_n \rangle$, ordered by their timestamps, i.e., $u_{i\mid t} < u_{j\mid t}$ for any $i,j such that 1 \leq i < j \leq n$.
\end{definition}

\begin{table}[tbh]
\makebox[0.98\textwidth][c]{
\centering
{\tiny{
\begin{tabular}{c|c|c|c|c|c|c|c}

\textbf{UI} & \textbf{UI} & \multicolumn{6}{|c}{\textbf{Parameter Names}}
\\\cline{3-8}
\textbf{Group} & \textbf{Type} & \textbf{P1} & \textbf{P2} & \textbf{P3} & \textbf{P4} & \textbf{P5} & \textbf{P6}\\\hline
\multirow{9}{*}{Navigation} & Create New Tab (Web) & ID & & &  & & \\
& Select Tab (Web) & URL & ID & Title &  & & \\
& Close Tab (Web) & URL & ID & Title &  & & \\
& Navigate To (Web) & URL & & &  & & \\
& Add Worksheet (Excel) & Workbook name & Worksheet name & &  & & \\
& Select Worksheet (Excel) & Workbook name & Worksheet name & &  & & \\
& Select Cell (Excel) & Workbook name & Worksheet name & Cell column & Cell row & Value \\
& Select Range (Excel) & Workbook name & Worksheet name & Range columns & Range rows & Value \\
& Select Field (Web) & URL & Name & ID & Value & & \\\hline
\multirow{3}{*}{Read} & Copy (Web) & URL & Name & ID & Value & Copied content \\
& Copy Cell (Excel) & Workbook name & Worksheet name & Cell column & Cell row & Value & Copied content \\
& Copy Range (Excel) & Workbook name & Worksheet name & Range columns & Range rows & Value & Copied content \\\hline
\multirow{9}{*}{Write} 
& Paste Into Cell (Excel) & Workbook name & Worksheet name & Cell column & Cell row & Value & Pasted content  \\
& Paste Into Range (Excel) & Workbook name & Worksheet name & Range columns & Range rows & Value & Pasted content \\
& Paste (Web) & URL & Name & ID & Value & Pasted content \\
& Click Button (Web) & URL & Name & ID & Type & & \\
& Click Link (Web) & URL & Inner text & Href &  & & \\
& Edit Field (Web) & URL & Name & ID & Type & Value & \\
& Edit Cell (Excel) & Workbook name & Worksheet name & Cell column & Cell row & Value & \\
& Edit Range (Excel) & Workbook name & Worksheet name & Range columns & Range rows & Value & \\\hline
\end{tabular}
}}}
\caption{User interaction types and their parameters}
\label{tab:uiParam}
\end{table}

Ideally, UIs recorded in a log should only relate to the execution of the task(s) of interest.
However, in practice, a log often also contains UIs that do not contribute to completing the recorded task(s).
We can consider such UIs to be \emph{noise}. 
Examples of noise UIs include a worker browsing the web (e.g., social networking) while executing a task that does not require to do that, or a worker committing mistakes (e.g., filling a text field with an incorrect value or copying a wrong cell of a spreadsheet). 
While we cannot detect the former kind of noise without a context-aware noise filter, we can identify the latter type of noise. 
Given that noise in a log may negatively affect the segmentation step, we attempt to remove it. 
Specifically, the filter we implemented removes UIs whose effects are overwritten by subsequent UIs, and certain navigation UIs that a software robot would not need to replicate. 
To identify and remove such UIs, we rely on three search-and-replace rules defined as regular expressions that operate as follows.

\begin{itemize}
\item[1.] Remove UIs of type \emph{select cell}, \emph{select range}, \emph{select field} (e.g., Table~\ref{tab:uiLog}, rows 2, 4, 7);
\item[2.] Remove UIs of type \emph{copy} that are not eventually followed by UI of type \emph{paste} before another UI of type \emph{copy} occurs (e.g., Table~\ref{tab:uiLog}, row 42);
\item[3.] Remove UIs of type \emph{edit cell}, \emph{edit range}, and \emph{edit field} that are followed by another UI of the same type that targets the same cell or field and overwrites its content before a UI of type \emph{copy} occurs (e.g., Table~\ref{tab:uiLog}, row 22).
\end{itemize}
We note that, given an unsegmented log, it is impossible to apply the third rule straightforward, as removing the first UI of type \emph{edit} (considered redundant) may be an error if the second UI of type \emph{edit} belongs to a successive task execution.
Therefore, we postpone the application of the third rule after the segmentation step. 
The filtering rules are applied recursively on the log until no more UIs are removed and the log is assumed to be free of \emph{detectable} noise.
Devising and applying more sophisticated noise filtering algorithms would probably benefit the approach presented in this study.
However, the design of such algorithms is outside the scope of this paper, and we leave it as possible future work.

After filtering the log, the vast majority of UIs are unique because they differ by their unique payload.
Note that even the UIs capturing the same action within the same task execution (or different task executions) would appear different.
To discover each task execution recorded in the log, we need to detect all the UIs that even having different payloads correspond to the same action within the same or different task execution(s).

Given a UI, its payload can be divided into \emph{data parameters} and \emph{context parameters}. 
The former store the data values used during the execution of tasks, e.g., the value of text fields or copied content. 
Consequently, \emph{data parameters} usually have different values in different task executions. 
In contrast, the latter capture the context in which UIs were performed, e.g., the application and the location within the application. 
Therefore, \emph{context parameters} of the same UI within a task are likely to have the same values across different task executions.
For example, the payload of a UI of type \emph{copy cell} has the following parameters (see also Table~\ref{tab:uiParam}):
\emph{workbook name} (the Excel file name);
\emph{worksheet name} (within the Excel file);
\emph{cell column} (i.e., the column of the cell in the worksheet that was selected for the UI);
\emph{cell row} (i.e., the row of the cell in the worksheet that was selected for the UI);
\emph{value} (i.e., current value of the cell selected for the UI);
\emph{copied content} (the content copied as the result of the UI).
Here, \emph{workbook name}, \emph{worksheet name}, \emph{cell column/row} are \emph{context parameters},
while \emph{copied content} and \emph{value} are \emph{data parameters}.
Different context parameters characterize different UI types.
For example, a UI of type \emph{click button} performed in a web browser has only these context parameters: \emph{URL}; \emph{name} (i.e., the label of the button); \emph{ID} (of the button, as an element in the HTML page); and \emph{type}.
Often, context parameters are determined by the type of UI. 
To reduce the chance of possible automated misinterpretations, we allow the user to configure the context parameters of various UI types manually.


To segment an input UI log, we rely on the context parameters of the UIs.
We call a UI whose payload has been reduced to its context parameters a \emph{normalized UI}.
\begin{definition}[\textbf{Normalized UI}]\label{def:nui}
Given a UI $u =  (t, \tau, P_{\tau}, Z, \phi)$, the UI $\bar{u} = (t, \tau, \bar{P_{\tau}}, \bar{Z}, \phi)$ is its normalized version, where $\bar{Z}$ contains only the values of the parameters in $\bar{P_{\tau}}$, where $\bar{P_{\tau}}$ is a set of context parameters.
\end{definition}
Two normalized UIs $u_1 = (t_1, \tau, \bar{P_{\tau}}, \bar{Z_1}, \phi_1)$ and $u_2 = (t_2, \tau, \bar{P_{\tau}}, \bar{Z_2}, \phi_2)$ are \emph{equivalent}, denoted by $u_1 = u_2$ iff $\forall p \in \bar{P_{\tau}} \Rightarrow \phi_1(p) = \phi_2(p)$.

A log in which all the UIs have been normalized is a \emph{normalized log}, and we refer to it with the notation $\bar{\Sigma} = \langle \bar{u_1}, \bar{u_2}, \dots, \bar{u_n} \rangle$. 
Table~\ref{tab:uiLog} and Table~\ref{tab:norm-uilog} show, respectively, a fragment of a log and its normalized version. 
Intuitively, in a normalized log, the chances that two executions of the same task have the same sequence (or set) of normalized UIs are high because they have only context parameters. 
We leverage such a characteristic of the normalized log to identify its segments (i.e., start and end of each executed task), and then the routine(s) within the segments.
\begingroup
\begin{table}[hbtp]
\makebox[0.98\textwidth][c]{
\centering
{\tiny{
    \begin{tabular}{c|c|c|c|c|c|c}
    & \textbf{UI}
    & \textbf{UI}
    & \multicolumn{4}{|c}{\textbf{Payload}}
    \\\cline{4-7}

    \textbf{Row}
    & \textbf{Timestamp}
    & \textbf{Type}
    & \textbf{$P_1$}
    & \textbf{$P_2$}
    & \textbf{$P_3$}
    & \textbf{$P_4$}\\\hline

   1 & 2019-03-03T19:02:18 & Click button (Web) & http://www.unimelb.edu.au & New Record & newRecord & button \\\hline
   2 & 2019-03-03T19:02:23 & Copy cell (Excel) & StudentRecords & Sheet1 & A & -- \\\hline 
   3 & 2019-03-03T19:02:26 & Paste (Web) & http://www.unimelb.edu.au & Full Name & name & -- \\\hline 
   4 & 2019-03-03T19:02:28 & Edit field (Web) & http://www.unimelb.edu.au & Full Name & name & text \\\hline
   5 & 2019-03-03T19:02:31 & Copy cell (Excel) & StudentRecords & Sheet1 & B & -- \\\hline  
   6 & 2019-03-03T19:02:37 & Paste (Web) & http://www.unimelb.edu.au & Date & date & -- \\\hline 
   7 & 2019-03-03T19:02:40 & Edit field (Web) & http://www.unimelb.edu.au & Date & date & text \\\hline
   8 & 2019-03-03T19:07:33 & Copy cell (Excel) & StudentRecords & Sheet1 & C & -- \\\hline 
   9 & 2019-03-03T19:07:40 & Paste (Web) & http://www.unimelb.edu.au & Phone & phone & -- \\\hline 
 10 & 2019-03-03T19:07:48 & Edit field (Web) & http://www.unimelb.edu.au & Phone & phone & text \\\hline
 11 & 2019-03-03T19:07:50 & Copy cell (Excel) & StudentRecords & Sheet1 & D & -- \\\hline 
 12 & 2019-03-03T19:07:55 & Paste (Web) & http://www.unimelb.edu.au & Country of residence & country & -- \\\hline 
 13 & 2019-03-03T19:08:02 & Edit field (Web) & http://www.unimelb.edu.au & Country of residence & country & text \\\hline
 14 & 2019-03-03T19:08:05 & Edit field (Web) & http://www.unimelb.edu.au & Student status & status & select \\\hline   
 15 & 2019-03-03T19:08:08 & Click button (Web) & http://www.unimelb.edu.au & Submit & submit & submit \\\hline
 16 & 2019-03-03T19:08:12 & Click button (Web) & http://www.unimelb.edu.au & New Record & newRecord & button \\\hline
 17 & 2019-03-03T19:08:17 & Copy cell (Excel) & StudentRecords & Sheet1 & B & -- \\\hline  
 18 & 2019-03-03T19:08:21 & Paste (Web) & http://www.unimelb.edu.au & Date & date & -- \\\hline 
 19 & 2019-03-03T19:08:28 & Edit field (Web) & http://www.unimelb.edu.au & Date & date & text \\\hline
 20 & 2019-03-03T19:08:35 & Copy cell (Excel) & StudentRecords & Sheet1 & C & -- \\\hline 
 21 & 2019-03-03T19:08:38 & Paste (Web) & http://www.unimelb.edu.au & Phone & phone & -- \\\hline 
 22 & 2019-03-03T19:08:39 & Edit field (Web) & http://www.unimelb.edu.au & Phone & phone & text \\\hline
 23 & 2019-03-03T19:08:40 & Copy cell (Excel) & StudentRecords & Sheet1 & A & -- \\\hline 
 24 & 2019-03-03T19:08:42 & Paste (Web) & http://www.unimelb.edu.au & Full Name & name & -- \\\hline 
 25 & 2019-03-03T19:08:43 & Edit field (Web) & http://www.unimelb.edu.au & Full Name & name & text \\\hline
 26 & 2019-03-03T19:08:45 & Copy cell (Excel) & StudentRecords & Sheet1 & D & -- \\\hline 
 27 & 2019-03-03T19:08:47 & Paste (Web) & http://www.unimelb.edu.au & Country of residence & country & -- \\\hline 
 28 & 2019-03-03T19:08:49 & Edit field (Web) & http://www.unimelb.edu.au & Country of residence & country & text \\\hline
 29 & 2019-03-03T19:08:52 & Edit field (Web) & http://www.unimelb.edu.au & Student status & status & select \\\hline   
 30 & 2019-03-03T19:08:53 & Click button (Web) & http://www.unimelb.edu.au & Submit & submit & submit \\\hline
 \dots & \dots & \dots & \dots & \dots & \dots & \dots \\\hline
 \end{tabular}
    }}}
   \caption{Normalized user interaction log after preprocessing}\label{tab:norm-uilog}
   \vspace{-2mm}
\end{table}
\endgroup

\subsection{Segmentation}
\label{sec:segmentation}


A log may capture long working sessions, where a worker performs multiple instances of one or more tasks. 
The next step of our approach decomposes the log into \emph{segments} that identify the start and the end of each recorded task in the log. 
Given a normalized log, we generate its control-flow graph (CFG). 
A CFG is a graph where each vertex represents a different normalized UI, and each edge captures a directly-follows relation between the two normalized UIs represented by the source and the target vertices of the edge. 
A CFG has an explicit source vertex representing the first normalized UI recorded in the log.

\begin{figure}[htb]
\centering
\subfloat[Before\label{fig:original}]{
\includegraphics[scale = 0.95]{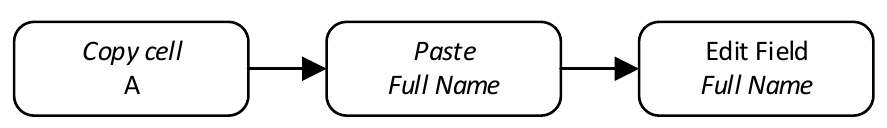}
}
\hspace{1cm}
\subfloat[After\label{fig:collapsed}]{
\includegraphics[scale = 0.85]{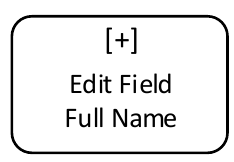}
}
\caption{Collapsed vertices in Figure~\ref{fig:cfg}}
\label{fig:collapsing}
\end{figure}

\begin{figure}[htb]
\centering
\hspace*{-2cm}\includegraphics[scale = 0.8]{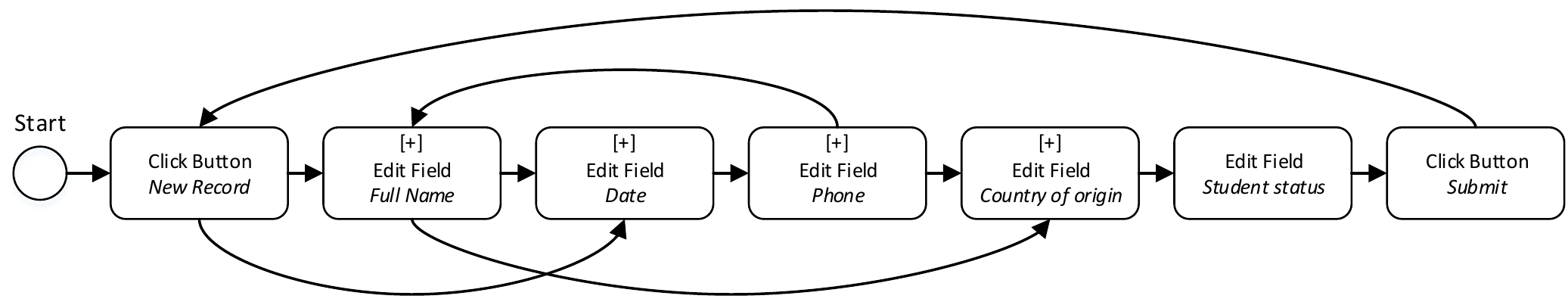}
\caption{Example of a Control-Flow Graph}
\label{fig:cfg}
\end{figure}

Given a log, the directly follows relation on UI is defined as follows.
\begin{definition}[\textbf{Directly-follows relation}]
Let $\bar{\Sigma} = \langle \bar{u}_1, \bar{u}_2, \dots, \bar{u}_n \rangle$ be a normalized log. Given two UIs, $\bar{u}_x, \bar{u}_y \in \bar{\Sigma}$, we say that $\bar{u}_y$ directly-follows $\bar{u}_x$, i.e., $\bar{u}_x \leadsto \bar{u}_y$, iff $\bar{u}_{x\mid t} < \bar{u}_{y\mid t} \wedge \nexists \bar{u}_z \in \bar{\Sigma} \mid \bar{u}_{x\mid t} \leq \bar{u}_{z\mid t} \leq \bar{u}_{y\mid t}$.
\end{definition}
\begin{definition}[\textbf{Control-Flow Graph (CFG)}]
Given a normalized log, $\bar{\Sigma} = \langle \bar{u_1}, \bar{u_2}, \dots, \bar{u_n} \rangle$, let $\bar{A}$ be the set of all the normalized UIs in $\bar{\Sigma}$. A Control-Flow Graph (CFG) is a tuple $G = (V, E, \hat{v}, \hat{e})$, where:
$V$ is the set of vertices of the graph, each vertex maps one UI in $\bar{A}$;
$E \subseteq V \times V$ is the set of edges of the graph, and each $(v_i, v_j) \in E$ represents a directly-follows relation between the UIs mapped by $v_i$ and $v_j$;
$\hat{v}$ is the graph \emph{entry vertex}, such that $\forall v \in V \nexists (v, \hat{v}) \in E \wedge \nexists (\hat{v}, v) \in E$;
while $\hat{e} = (\hat{v}, v_0)$ is the graph \emph{entry edge}, such that $v_0$ maps $\bar{u_1}$.
We note that $\hat{v} \notin V$, and $\hat{e} \notin E$, since they are artificial elements of the graph.
\end{definition}
It is likely that a CFG is cyclic, since a loop represents the start of a new execution of the task recorded in the log. Indeed, in an ideal scenario, once a task execution ends with a certain UI (a vertex in the CFG), the next UI (i.e., the first UI of the next task execution) should have already been mapped to a vertex of the CFG, and a loop will be generated.
In such a case, all the vertices in the loop represent the UIs performed during the execution of the task.
If several different tasks are recorded in sequence in the same log, we would observe several disjoint loops in the CFG, while if a task has repetitive subtasks, we would observe nested loops in the CFG. 
\figurename~\ref{fig:cfg} shows the CFG generated from the log captured in Table~\ref{tab:norm-uilog}, we note that for simplicity we collapsed some vertices as shown in Figure~\ref{fig:collapsing}.

Once the CFG is generated, we turn our attention to identifying its back-edges (i.e., its loops). By identifying the CFG back-edges and their UIs, we extract the start and end UIs of the repeated task. These UIs are used to mark the boundaries between task executions. The back-edges of a CFG can be identified by analyzing the CFG Strongly Connected Components (SCCs). Given a graph, an SCC is a subgraph where for all its pairs of vertices, there exist a set of edges connecting the pair of vertices such that all the sources and targets of these edges belong to the subgraph.
\begin{definition}[\textbf{CFG Path}]
Given a CFG $G = (V, E, \hat{v}, \hat{e})$, a CFG path is a sequence of vertices $p_{v_1,v_k} = \langle v_1, \dots, v_k \rangle$ such that for each $i \in [1,k-1] \Rightarrow v_i \in V \cup \{ \hat{v} \} \wedge \exists (v_i, v_{i+1}) \in E \cup \{ \hat{e}\}$.
\end{definition}
\begin{definition}[\textbf{Strongly Connected Component (SCC)}]
Given a graph $G = (V, E, \hat{v}, \hat{e})$, a strongly connected component (SCC) of G is a pair $\delta = (\bar{V}, \bar{E})$, where $\bar{V} = \{ v_1, v_2, \dots, v_m \} \subseteq V$ and $\bar{E} = \{ e_1, e_2, \dots, e_k \} \subseteq E$ such that $\forall v_i, v_j \in \bar{V} \exists p_{v_i,v_j} \mid \forall v \in p \Rightarrow v \in \bar{V}$. Given an SCC $\delta = (\bar{V}, \bar{E})$, we say that $\delta$ is \emph{non-trivial} iff $\left| \bar{V} \right| > 1$. Given a graph $G$, $\Delta_G$ denotes the set of all the non-trivial SCCs in G.
\end{definition}

\begin{figure}[htb]
\centering
\includegraphics[scale = 0.9]{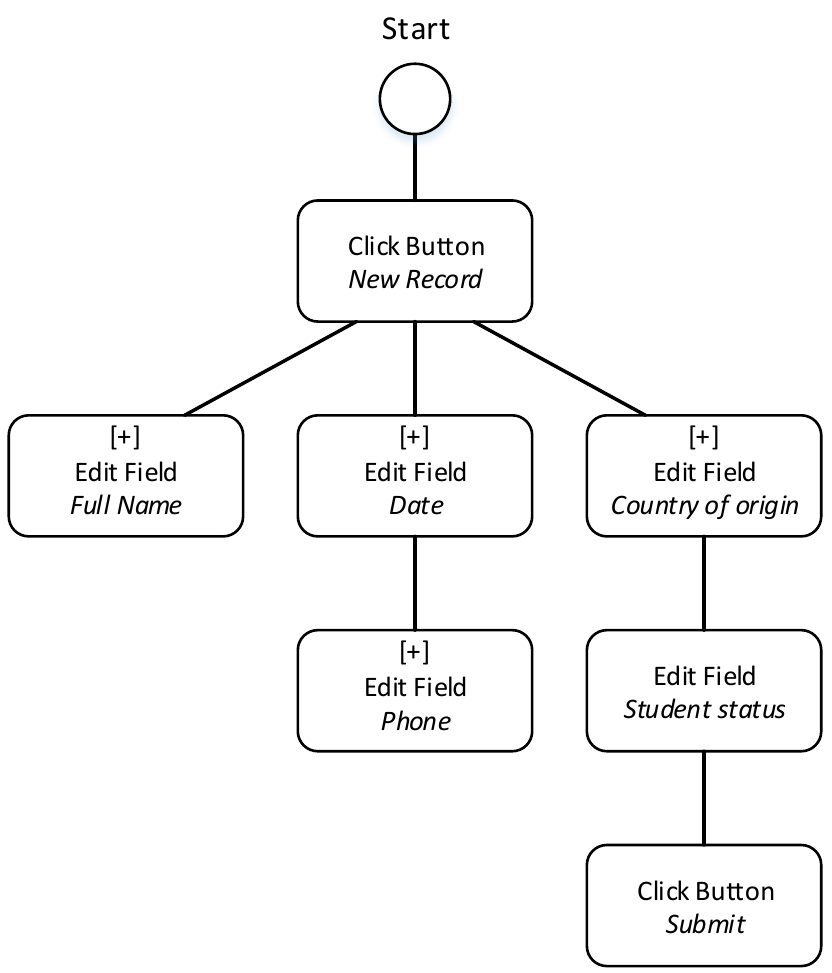}
\caption{Dominator tree}
\label{fig:domTree}
\end{figure}

Algorithm~\ref{alg:beDetection} and Algorithm~\ref{alg:analyseSCC} describe how we identify the SCCs of the CFG. Given a CFG $G = (V,E,\hat{v},\hat{e})$, we first build its dominator tree $\Theta$ (Algorithm~\ref{alg:beDetection}, line~\ref{alg:domTree}), which captures domination relations between the vertices of the CFG. \figurename~\ref{fig:domTree} shows the dominator tree of the CFG in \figurename~\ref{fig:cfg}.
Then, we discover the set of all non-trivial SCCs ($\Delta_G$) by applying the Kosaraju's algorithm \cite{sharir1981strong} and removing the trivial SCCs (Algorithm~\ref{alg:beDetection}, line~\ref{alg:scc}). For each $\delta = (\bar{V}, \bar{E}) \in \Delta_G$, we discover its \emph{header} using the dominator tree (Algorithm~\ref{alg:analyseSCC}, line~\ref{alg:header}). The header of a dominator tree $\delta$ is a special vertex $\hat{h} \in \bar{V}$, such that $\forall p_{\hat{v},v} \mid v \in \bar{V} \Rightarrow \hat{h} \in p_{\hat{v},v}$, i.e., the \emph{header} $\hat{h}$ (a.k.a. the SCC entry) is the SCC vertex that dominates all the other SCC vertices. Once we have $\hat{h}$, we can identify the back-edges as $(v,\hat{h})$ with $v \in \bar{V}$ (line~\ref{alg:incoming}). Finally, the identified back-edges are stored and removed (lines~\ref{alg:backEdges} and ~\ref{alg:edgesSub}) in order to look for nested SCCs and their back-edges by recursively executing Algorithm~\ref{alg:analyseSCC} (line~\ref{alg:recursion}), until no more SCCs and back-edges are found. However, if we detect an SCC that does not have a header vertex (formally, the SCC is irreducible), we cannot identify the SCC back-edges. In such a case, we collect via a depth-first search of the CFG the edges $(v_x, v_y) \in \bar{E}$ such that $v_y$ is topologically deeper than $v_x$ - we call these edges \emph{loop-edges} of the SCC (line~\ref{alg:loops}). Then, out of all the loop-edges, we store (and remove from the SCC) the one having target and source connected by the longest \emph{simple path} entirely contained within the SCC (lines~\ref{alg:deepestEdge} to ~\ref{alg:removeEdge}).

Given the CFG presented in \figurename~\ref{fig:cfg} and its corresponding dominator tree (see \figurename~\ref{fig:domTree}), we identify the SCC that consists of all the vertices except the \emph{entry vertex}. Then, by applying Algorithm~\ref{alg:analyseSCC}, we identify: the SCC header -- \emph{Click Button [New Record]}; and the only back-edge -- (\emph{Click Button [Submit]}, \emph{Click Button [New Record]}), which we save and remove from the SCC. After the removal of this back-edge, we identify the nested SCC that contains edits of \emph{Full Name}, \emph{Date}, and \emph{Phone} fields. Note that this second SCC does not have a header because it is irreducible, due to its multiple entries (\emph{Edit Field [Full Name]} and \emph{Edit Field [Date]}). However, by applying the depth-first search, we identify as candidate loop-edge for removal: (\emph{Edit Field [Phone]}, \emph{Edit Field [Full Name]}). After we remove this edge from the CFG, no SCCs are left, so Algorithm~\ref{alg:analyseSCC} terminates.

\begin{algorithm}[btp]
{   \scriptsize
	\Input{CFG $G$}
	\Output{Back-edges Set $B$}
	\BlankLine
	
	$B \leftarrow \varnothing$\;
	Dominator Tree $\Theta \leftarrow$ computeDominatorTree($G$)\;\label{alg:domTree}
	Set $\Delta_G \leftarrow$ findSCCs($G$)\; \label{alg:scc}
	
	\lForEach{$\delta \in \Delta_G$}{
		AnalyseSCC($\delta$, $\Theta$, $B$) \label{alg:analyse}
	}
	\BlankLine
	\Return $B$\;
	
	\caption{Back-edges detection}\label{alg:beDetection}
}
\end{algorithm}

\begin{algorithm}[btp]
{
    \scriptsize
	\Input{SCC $\delta = (\bar{V}, \bar{E})$, Dominator Tree $\Theta$, Back-edges Set $B$}
	\BlankLine

		Header $\hat{h} \leftarrow$ findHeader($\delta$, $\Theta$)\; \label{alg:header}
		\eIf{$\hat{h} \neq$ null}{
			Set $I \leftarrow$ getIncomingEdges($\delta$, $\hat{h}$)\; \label{alg:incoming}
			$B \leftarrow B \cup I$\;  \label{alg:backEdges}
			$\bar{E} \leftarrow \bar{E} \setminus I$\; \label{alg:edgesSub}
		}{
			Set $L \leftarrow$ findLoopEdges($\delta$)\; \label{alg:loops}
			Edge $e \leftarrow$ getTheDeepestEdge($\delta$, $L$)\; \label{alg:deepestEdge}
			\textbf{remove} $e$ \textbf{from} $\bar{E}$\; \label{alg:removeEdge}
		}
		Set $\Delta_\delta \leftarrow$ findSCCs($\delta$)\; \label{alg:findSCCs}
		\lForEach{$\gamma \in \Delta_\delta$}{
			AnalyseSCC($\gamma$, $\Theta$, $B$) \label{alg:recursion}
		}
	
	\caption{Analyse SCC}\label{alg:analyseSCC}
}
\end{algorithm} 

At this point, we collected all the back-edges of the CFG. Next, we use them to segment the log. We do so by applying Algorithm~\ref{alg:segIdentification}. First, we retrieve all the targets and sources of all the back-edges in the CFG and collect their corresponding UIs (lines~\ref{alg:targets4} and~\ref{alg:sources4}). Each UI mapped onto a back-edge target is an eligible segment starting point (from now on, \emph{segment-start UI}). A back-edge conceptually captures the end of a task execution, while its target represents the first UI of the next task execution. By applying the same reasoning, each UI mapped onto the source of a back-edge is an eligible segment ending point (hereinafter, \emph{segment-end UI}). Then, we sequentially scan all the UIs in the log (line~\ref{alg:uilogscan4}). When we encounter a segment-start UI (line~\ref{alg:segstart4}), and we are not already within a segment (see line~\ref{alg:notinsegment4}), we create a new segment ($s$, a list of UIs), we append the segment-start UI ($\bar{u}$), and we store it in order to match it with the correct segment-end UI (line~\ref{alg:startsegment41} to~\ref{alg:startsegment42}). Our strategy to detect segments in the log is driven by the following underlying assumption: a specific segment-end UI will be followed by the same segment-start UI so that we can match segment-end and segment-start UIs exploiting back-edge's sources and targets (respectively). If the UI is not a segment-start (line~\ref{alg:nostart4}), we check if we are within a segment (line~\ref{alg:insegment4}) and,
if not, we discard the UI, assuming it is noise since it fell between the previous segment-end UI and the next segment-start UI. Otherwise, we append the UI to the current segment, and we check if this UI is a segment-end matching the current segment-start UI (line~\ref{alg:startendmatching4}).
If that is the case, we reached the end of the segment, and we add it to the set of segments (line~\ref{alg:segmentcomplete4}); otherwise, we continue reading the segment.

\newpage

\begin{algorithm}[tbp]
{
    \scriptsize
	\Input{Normalized UI log $\bar{\Sigma}$, Back-edges Set $B$}
	\Output{Segments List $\Psi$}
	\BlankLine
	
	Set $\Psi \leftarrow \varnothing$\;
	Set $T \leftarrow$ getTargets($B$)\;\label{alg:targets4}
	Set $S \leftarrow$ getSources($B$)\;\label{alg:sources4}
	Boolean WithinSegment $ \leftarrow$ false\;
	Normalized UI $u_0 \leftarrow$ null\;
	Queue $s \leftarrow \varnothing$\;
	
	\BlankLine
	\For{$i \gets 1$ \KwTo size$\left( \nuilog \right)$}{\label{alg:uilogscan4}
		Normalized UI $\nui \leftarrow$ getUI($\nuilog$, $i$)\;
		\eIf{$\nui \in T$}{\label{alg:segstart4}
			\eIf{WithinSegment $ = false$}{\label{alg:notinsegment4}
				$s \leftarrow \varnothing$\;\label{alg:startsegment41}
				\textbf{append} $\nui$ \textbf{to} $s$\;
				$u_0 \leftarrow \nui$\;
				WithinSegment $ \leftarrow$ true\;\label{alg:startsegment42}
			}{
				\textbf{append} $\nui$ \textbf{to} $s$\;
			}
 		}{\label{alg:nostart4}
 			\If{WithinSegment $= true$}{\label{alg:insegment4}
				\textbf{append} $\nui$ \textbf{to} $s$\;
				\If{$\nui \in S \wedge (\nui, u_0) \in B$}{\label{alg:startendmatching4}
					$\Psi \leftarrow \Psi \cup \{ s \}$\;\label{alg:segmentcomplete4}
					WithinSegment $ \leftarrow false$\;
				}
			}	
		}
	}
	\BlankLine
	\Return $\Psi$\;
	
	\caption{Segmentation}\label{alg:segIdentification}
}
\end{algorithm} 

Table~\ref{tab:segments} shows the segment-start and the segment-end UIs (highlighted in green and red, respectively), which delimits two segments within the normalized UI log of our running example (see also Table~\ref{tab:norm-uilog}).

\begingroup
\begin{table}[hbtp]
\makebox[0.98\textwidth][c]{
\centering
{\tiny{
    \begin{tabular}{c|c|c|c|c|c|c}
    & \textbf{UI}
    & \textbf{UI}
    & \multicolumn{4}{|c}{\textbf{Payload}}
    \\\cline{4-7}

    \textbf{Row}
    & \textbf{Timestamp}
    & \textbf{Type}
    & \textbf{$P_1$}
    & \textbf{$P_2$}
    & \textbf{$P_3$}
    & \textbf{$P_4$}\\\hline

   \rowcolor{green}
   1 & 2019-03-03T19:02:18 & Click button (Web) & http://www.unimelb.edu.au & New Record & newRecord & button \\\hline
   2 & 2019-03-03T19:02:23 & Copy cell (Excel) & StudentRecords & Sheet1 & A & -- \\\hline 
   3 & 2019-03-03T19:02:26 & Paste (Web) & http://www.unimelb.edu.au & Full Name & name & -- \\\hline 
   4 & 2019-03-03T19:02:28 & Edit field (Web) & http://www.unimelb.edu.au & Full Name & name & text \\\hline
   5 & 2019-03-03T19:02:31 & Copy cell (Excel) & StudentRecords & Sheet1 & B & -- \\\hline  
   6 & 2019-03-03T19:02:37 & Paste (Web) & http://www.unimelb.edu.au & Date & date & -- \\\hline 
   7 & 2019-03-03T19:02:40 & Edit field (Web) & http://www.unimelb.edu.au & Date & date & text \\\hline
   8 & 2019-03-03T19:07:33 & Copy cell (Excel) & StudentRecords & Sheet1 & C & -- \\\hline 
   9 & 2019-03-03T19:07:40 & Paste (Web) & http://www.unimelb.edu.au & Phone & phone & -- \\\hline 
 10 & 2019-03-03T19:07:48 & Edit field (Web) & http://www.unimelb.edu.au & Phone & phone & text \\\hline
 11 & 2019-03-03T19:07:50 & Copy cell (Excel) & StudentRecords & Sheet1 & D & -- \\\hline 
 12 & 2019-03-03T19:07:55 & Paste (Web) & http://www.unimelb.edu.au & Country of residence & country & -- \\\hline 
 13 & 2019-03-03T19:08:02 & Edit field (Web) & http://www.unimelb.edu.au & Country of residence & country & text \\\hline
 14 & 2019-03-03T19:08:05 & Edit field (Web) & http://www.unimelb.edu.au & Student status & status & select \\\hline   
 \rowcolor{red}  
 15 & 2019-03-03T19:08:08 & Click button (Web) & http://www.unimelb.edu.au & Submit & submit & submit \\\hline
 \rowcolor{green}
 16 & 2019-03-03T19:08:12 & Click button (Web) & http://www.unimelb.edu.au & New Record & newRecord & button \\\hline
 17 & 2019-03-03T19:08:17 & Copy cell (Excel) & StudentRecords & Sheet1 & B & -- \\\hline  
 18 & 2019-03-03T19:08:21 & Paste (Web) & http://www.unimelb.edu.au & Date & date & -- \\\hline 
 19 & 2019-03-03T19:08:28 & Edit field (Web) & http://www.unimelb.edu.au & Date & date & text \\\hline
 20 & 2019-03-03T19:08:35 & Copy cell (Excel) & StudentRecords & Sheet1 & C & -- \\\hline 
 21 & 2019-03-03T19:08:38 & Paste (Web) & http://www.unimelb.edu.au & Phone & phone & -- \\\hline 
 22 & 2019-03-03T19:08:39 & Edit field (Web) & http://www.unimelb.edu.au & Phone & phone & text \\\hline
 23 & 2019-03-03T19:08:40 & Copy cell (Excel) & StudentRecords & Sheet1 & A & -- \\\hline 
 24 & 2019-03-03T19:08:42 & Paste (Web) & http://www.unimelb.edu.au & Full Name & name & -- \\\hline 
 25 & 2019-03-03T19:08:43 & Edit field (Web) & http://www.unimelb.edu.au & Full Name & name & text \\\hline
 26 & 2019-03-03T19:08:45 & Copy cell (Excel) & StudentRecords & Sheet1 & D & -- \\\hline 
 27 & 2019-03-03T19:08:47 & Paste (Web) & http://www.unimelb.edu.au & Country of residence & country & -- \\\hline 
 28 & 2019-03-03T19:08:49 & Edit field (Web) & http://www.unimelb.edu.au & Country of residence & country & text \\\hline
 29 & 2019-03-03T19:08:52 & Edit field (Web) & http://www.unimelb.edu.au & Student status & status & select \\\hline   
 \rowcolor{red}
 30 & 2019-03-03T19:08:53 & Click button (Web) & http://www.unimelb.edu.au & Submit & submit & submit \\\hline
 \dots & \dots & \dots & \dots & \dots & \dots & \dots \\\hline
 \end{tabular}
    }}}
   \caption{Segments identification}\label{tab:segments}
   \vspace{-2mm}
\end{table}
\endgroup

\subsection{Candidates routines identification}
\label{sec:candidatesDiscovery}

Once the log has been segmented, we move to the identification of the candidate routines. The identification step is based on the CloFast sequence mining algorithm \cite{fumarola2016clofast}. To integrate CloFast in our approach, we have to define the structure of the sequential patterns we want to identify. In this paper, we define a \emph{sequential pattern} within a UI log as a sequence of normalized UIs always occurring in the same order in different segments, yet allowing gaps between the UIs belonging to the pattern. For example, if we consider the following three segments:
$\langle u_1, u_y, u_2, u_3 \rangle$,
$\langle u_1, u_2, u_x, u_3 \rangle$,
and $\langle u_1, u_x, u_2, u_3 \rangle$;
they all contain the same sequential pattern that is $\langle u_1, u_2, u_3 \rangle$.

Furthermore, we define the \emph{support} of a sequential pattern as the ratio of segments containing the pattern and the total number of segments. 
We refer to \emph{closed} patterns and \emph{frequent} patterns (relatively to an input threshold) as they are known in the literature. Specifically, a frequent pattern is a pattern that appears in at least a number of occurrences indicated by the threshold, while a closed pattern is a pattern that is not included in another pattern having exactly the same support. By applying CloFast to the log segments, we discover all the \emph{frequent closed} sequential patterns.

Some of these patterns may be \emph{overlapping}, which (in our context) means that they share some UIs. An example of overlapping patterns is the following, given three segments:
$\langle u_1, u_y, u_2, u_3, u_x, u_4 \rangle$,
$\langle u_1, u_y, u_2, u_x, u_3, u_4 \rangle$,
and $\langle u_1, u_x, u_2, u_3, u_4 \rangle$;
$\langle u_1, u_2, u_3, u_4 \rangle$ and $\langle u_1, u_x, u_4 \rangle$ are sequential patterns, but they overlap due to the shared UIs: $u_1$ and $u_4$. In practice, each UI belongs to only one routine, therefore, we are interested in discovering only non-overlapping patterns. For this purpose, we implemented an optimization that we use on top of CloFast. Given the set of patterns discovered by CloFast, we rank them by a pattern quality criterion, and we select the best pattern (i.e., the top one in the ranking). We integrated four pattern quality criteria to select the candidate routines: pattern frequency, pattern length, pattern coverage, and pattern cohesion score~\cite{DBLP:conf/iui/DevL17}. Pattern frequency considers how many times the pattern was observed in different segments.
Pattern length considers the length of the patterns. Pattern coverage considers the percentage of the log that is covered by all the pattern occurrences. 
Finally, pattern cohesion score considers the level of adjacency of the elements inside a pattern. It is calculated as the difference between the pattern length and the median number of gaps between its elements. In other words, cohesion prioritizes the patterns whose UIs appear consecutively without (or with few) gaps while taking into account also the pattern length.

For the candidate routine that we identified as the best pattern for a given quality criterion, we collect and remove all its occurrences from the log. An occurrence of a candidate routine is called a \emph{routine instance}. Formally, a routine instance is a sequence of (non-normalized) UIs, e.g., $r = \langle u_1, u_2, u_3, u_4 \rangle$. After the removal of all the instances of the best candidate routine from the log,
we repeat this identification step until no more candidate routines are identified. At the completion of this step, we obtain a set of candidate routines, referred to as $\crset$, such that, for each candidate routine $c_i \in \crset$, we can retrieve the set of its routine instances, referred to as $\riset$.

Considering our running example, with reference to Table~\ref{tab:segments}, assuming that the two routine instances that we identified in the previous step (by detecting their segment-start and segment-end UIs) frequently occur in the original log (a snapshot of which is captured in Table~\ref{tab:uiLog}), and choosing length as a selection criterion, at the end of this step, we would discover two candidate routines, each consisting of 15 normalized UIs (as shown in Table~\ref{tab:segments}). An example of a routine instance for each of the two candidate routines can be easily observed in the original log, Table~\ref{tab:uiLog} rows 1 to 24 and 25 to 49 (excluding the UIs filtered in the first step of our approach). 


\subsection{Automatability assessment}
\label{sec:automatabilityAssessment}

The candidate routines in $\crset$ (and their instances, $\riset$) that we identified in the previous step represent behavior recorded in the log that frequently repeats itself, thus it is the candidate for automation. However, the fact that a routine is frequently observed in a log is not a sufficient condition to guarantee its automatability. Let us consider the following example; a worker fills in and submits 100 times the same web-form, doing it always with the same sequence of actions but inputting manually-generated data (e.g., received over a phone call or copied from a hard-copy document). In such a scenario, although we would identify the filling and submission of the web-form as a candidate routine, we would not be able to automate it because we cannot automatically generate the data in input to the web-forms. On the other hand, if the data in input to the web-forms was copied from another digital document, for example a spreadsheet, we could probably automate the routine.

Considering such a context, the next step of our approach is to assess the degree of automatability of the discovered candidate routines. To do so, given a candidate routine $c_i \in \crset$, we check whether all its UIs are deterministic. We consider a UI to be deterministic if a software robot can replicate its execution. This is possible when: i) the input data of a UI can be determined automatically; or ii) the input data of a UI can be provided as input by the user when deploying the software robot. According to such constraints, we can provide the following rules to check whether a UI is deterministic or not.

\begin{itemize}
\item[1.] UIs belonging to the \emph{navigation} group (see Table~\ref{tab:uiParam}) are always deterministic because they do not take in input any data; except the \emph{select cell}, \emph{select field}, and \emph{select range} UIs which are removed during the filtering of the log (as described in Section~\ref{sec:segmentation});

\item[2.] UIs belonging to the \emph{read} group are always deterministic because the only input they require is the source of the copied content (e.g., row and column of a cell), which is either constant or can be inputted by the user when deploying the software robot in UiPath; 

\item[3.] UIs belonging to the \emph{write} group that are of type \emph{click} are always deterministic because they do not take in input any data, except the information regarding the element to be clicked which is always constant for a given candidate routine (by construction);

\item[4.] UIs belonging to the \emph{write} group that are of type \emph{paste} are always deterministic because they always retrieve data from the same source (i.e., the system clipboard).

\item[5.] UIs belonging to the \emph{write} group that are of type \emph{edit} are the only ones that are not always deterministic. In fact, these UIs are deterministic only if it is possible to determine the updated value of the edited elements (e.g., the value of a cell in a spreadsheet or of a text field in the web browser after the UI is executed). Furthermore, it has also to be possible to determine the target of the editing, although this is usually constant (if a web element) or can be inputted by the user when deploying the software robot in UiPath.
\end{itemize}

Algorithm~\ref{alg:automatabilityAssessment} shows how we check these five rules given as input a candidate routine $c_i$ and its routine instances $\riset$, and how we compose the corresponding routine specification of the input $c_i$. The algorithm starts by initializing the set $E$ as a collection of \emph{edit} UI types (\emph{edit cell}, \emph{edit range}, \emph{edit field}). Then, it iterates over all the normalized UIs in the input $c_i$ by checking their types. If the type of a normalized UI $\nui$ is not in $E$ (line~\ref{alg:checktype}), i.e., one of the rules 1 to 4 applies, we add it to the queue $D$, which stores all the deterministic UIs we identified. Otherwise, rule 5 applies. While rules 1 to 4 are simple checks on the UI types, the complexity of rule 5 required us to operationalize it through a separate algorithm, i.e., Algorithm~\ref{alg:checkeditUIs}, which is called within Algorithm~\ref{alg:automatabilityAssessment} (line~\ref{alg:calleditcheck}). Algorithm~\ref{alg:checkeditUIs} returns a pair $(d, \lambda)$, where $d$ is a \emph{boolean} (true if the input normalized UI is deterministic), and $\lambda$ is a \emph{data transformation step} required to automate $\nui$ and therefore available only if $\nui$ is deterministic. Once all the normalized UIs in the input $c_i$ have been checked, Algorithm~\ref{alg:automatabilityAssessment} outputs the \emph{routine specification} of $c_i$, as the pair ($c_i$, $\Lambda$), where $\Lambda$ is the set of all the \emph{data transformation steps} we collected by executing Algorithm~\ref{alg:checkeditUIs} (line~\ref{alg:calleditcheck}).

\begin{algorithm}[btp]
{
\scriptsize
	\Input{Candidate Routine $c_i$, Routine Instances Set $\riset$}
	\Output{Routine Specification ($c_i$, $\Lambda$)}
	\BlankLine

	Set $\Lambda \leftarrow \varnothing$\;\label{alg:setLambda}
	Set $E \leftarrow \{$ ``edit cell'', ``edit range'', ``edit field'' $\}$\;\label{alg:setE}
	Queue $D \leftarrow \varnothing$\;\label{alg:queueE}
	\BlankLine

	\ForEach{Normalized UI $\nui \in c_i$}{\label{alg:scanci}
		\eIf{getType($\nui$) $ \notin E$}{
			\textbf{append} $\nui$ \textbf{to} $D$\;\label{alg:checktype}
		}{
			$k \leftarrow $ checkUIofTypeEdit($\nui$, $c_i$, $\riset$)\;\label{alg:calleditcheck}
			Boolean $d \leftarrow $ getDeterministic($k$)\;\label{alg:istrue}
			\If{$d = $ true}{
				\textbf{append} $\nui$ \textbf{to} $D$\;
				$\Lambda \leftarrow \Lambda \ \cup $ getTransformationStep($k$)\;\label{alg:gettransform}
			}
		}
	}
	\BlankLine
	
	\Return ($c_i$, $\Lambda$)
	\caption{Routine automatability assessment}\label{alg:automatabilityAssessment}
}
\end{algorithm}
\begin{algorithm}[btp]
{
\scriptsize
	\Input{Normalized UI $\nui$, Candidate Routine $c_i$, Routine Instances Set $\riset$}
	\Output{Boolean $d$, Transformation Step $\lambda$}
	\BlankLine

	Boolean $d \leftarrow $ false\;\label{alg:setDeterminism}
          Set $C \leftarrow \{$ ``copy cell'', ``copy range'', ``copy field'' $\}$\;\label{alg:setC}
	Set $E \leftarrow \{$ ``edit cell'', ``edit range'', ``edit field'' $\}$\;\label{alg:setE}
           Set $P \leftarrow \{$ ``paste into cell'', ``paste into range'', ``paste'' $\}$\;\label{alg:setP}
	Set $T \leftarrow \varnothing$\;  
	Set $\Pi \leftarrow \varnothing$\;  
	Set $K \leftarrow \varnothing$\;  \label{alg:data2}
	Integer $n \leftarrow $ getPosition($\nui$, $c_i$)\;\label{alg:getPosition}
	\BlankLine
	
	\ForEach{$r \in \riset$}{\label{alg:traverseRoutineInstances}
		UI $u_1 \leftarrow $ get(r, n)\;\label{alg:getInstance}
		$K \leftarrow K \cup \{ u_1 \}$\;\label{alg:addInstance}
		$t_1 \leftarrow $ getTargetElement($u_1$)\;\label{alg:getTarget}
		$o \leftarrow $ getParameterValue($u_1$, ``Value'')\;\label{alg:getOutput}
		Queue $S \leftarrow \varnothing$\;\label{alg:queueS}
		Queue $I \leftarrow \varnothing$\;\label{alg:queueI}
		\For{$i \gets n$ \KwTo 1}{\label{alg:startMainIteration}
			UI $u_2 \leftarrow $ get(r, i)\;\label{alg:getU2}
			$\Pi \leftarrow \Pi \cup \{ (r, u_2) \}$\;\label{alg:addU2}

			\eIf{getType($u_2$) $ \in P$}{\label{alg:startPasteCheck}
				$t_2 \leftarrow $ getTargetElement($u_2$)\;
					\If{$t_2 = t_1$} {
						\For{$j \gets  i$ \KwTo 1}{
							UI $u_3 \leftarrow $ get(r, j)\;\label{alg:getU3}
							\If{getType($u_3$) $ \in C$}{\label{alg:checkForCopy}
								$s \leftarrow $ getTargetElement($u_3$)\;
								\textbf{append} $s$ \textbf{to} $S$\;\label{alg:addSource1}
								\textbf{append} getParameterValue($u_3$, ``Value'') \textbf{to} $I$\;\label{alg:addInput1}
								\Break
							}
						}
					}\label{alg:stopPasteCheck}
			}{
				\If{getType($u_2$) $ \in E$}{\label{alg:checkForEdit}
					$t_2 \leftarrow $ getTargetElement($u_2$)\;
					\If{$t_2 = t_1$} {\label{alg:sameEditTarget}
						\textbf{push} $t_2$ \textbf{to} $S$\;\label{alg:addSource2}
						\textbf{push} getParameterValue($u_2$,``Value'') \textbf{to} $I$\;\label{alg:addInput2}
						\Break
					}
				}
			}
		}\label{alg:stopMainIteration}
		$T \leftarrow T \cup \{ (I, o) \}$\;\label{alg:addTransformationExample}
	}

	Transformation $\chi \leftarrow $ discoverTransformation($T$)\;\label{alg:syntTransDiscoveryStart}
	\eIf{$\chi \neq null$}{
		$d \leftarrow $ true\;
		$\lambda \leftarrow $ ($S$, $\mathit{target}$, $\chi$)\;
	}{\label{alg:syntTransDiscoveryEnd}
		Set $D \leftarrow $ discoverDependencies($K$, $\Pi$)\;\label{alg:semTransDiscoveryStart}
		\If{$D \neq \varnothing$}{\label{alg:depfound}
			$d \leftarrow $ true\;\label{alg:tane1}
			$S \leftarrow $ getSources($D$)\;
			$\chi \leftarrow $ extractTransformation($D$)\;
			$\lambda \leftarrow $ ($S$, $\mathit{target}$, $\chi$)\;\label{alg:tane2}
		}\label{alg:semTransDiscoveryEnd}
	}
	\BlankLine
	
	\Return $(d, \lambda)$
	\caption{Check UI of Type Edit}\label{alg:checkeditUIs}
}
\end{algorithm}
Before moving to the final step of our approach, we describe how Algorithm~\ref{alg:checkeditUIs} verifies whether an input (normalized) UI of type \emph{edit} ($\nui$) is deterministic. In essence, Algorithm~\ref{alg:checkeditUIs} checks whether the value of the element edited by the execution of $\nui$ can be deterministically computed from the UIs observed before $\nui$ (in all the routine instances in $\riset$).
To do so, the algorithm looks for a possible data transformation function to compute the value of the edited element from the payloads of the UIs observed before $\nui$. If such a data transformation function exists, $\nui$ is considered to be deterministic, and the algorithm returns the identified function in the form of a data transformation step (which also includes source(s) and target of the data transformation function).
In the following, we walk through Algorithm~\ref{alg:checkeditUIs}.

We start by assuming that the UI in input is not deterministic, and we try to prove the opposite. We initialize to false the boolean variable which we will output at the end of the algorithm (line~\ref{alg:setDeterminism}), and we create the necessary data structures (line~\ref{alg:setC} to~\ref{alg:data2}). Given the input candidate routine $c_i$ and the normalized UI $\nui$, we extract the index of $\nui$ within $c_i$ (line~\ref{alg:getPosition}). Then, for each routine instance $r \in \riset$, we do what follows. 

We get the instance of the normalized UI $\nui$\footnote{We recall that a UI instance contains all the parameters, both context and data ones.} by retrieving the UI of index $n$ from $r$ (line~\ref{alg:getInstance}), and we store this UI ($u_1$) in the set $K$ (line~\ref{alg:addInstance}). We read the payload of $u_1$ to retrieve the target element ($t_1$, line~\ref{alg:getTarget}), $t_1$ can be the ID of a web browser element or the location of a cell in a spreadsheet. Also, we read the payload of $u_1$ to retrieve the value of the target element after the editing ($o$, line~\ref{alg:getOutput}). We initialize two queues, $S$ (which stands for \emph{sources}) and $I$ (which stands for \emph{inputs}). Queue $S$ stores the ID or location of the (source) element(s) that produced the data used by the \emph{edit} UI instance $u_1$; while queue $I$ stores the data that was used by the \emph{edit} UI instance $u_1$.

After this initialization, we iterate over all the UI instances preceding $u_1$ in $r$. Such an iteration goes backward from $u_1$ (position $n$ in $r$) till the first UI instance in $r$ (position 1) -- line~\ref{alg:startMainIteration} to~\ref{alg:stopMainIteration}, unless we identify another UI instance of type \emph{edit} performed on the same target element $t_1$ (see lines~\ref{alg:checkForEdit} to \ref{alg:sameEditTarget}). In the iteration captured between line~\ref{alg:startMainIteration} to~\ref{alg:stopMainIteration}, we do the following.

We store all the preceding UI instances ($u_2$) into the set $\Pi$, alongside the routine instance they belong to (i.e., we store a pair $(r, u_2)$ in $\Pi$). For each encountered $u_2$ of type \emph{paste}, we check its target element and we compare it to the target element of $u_1$. If they are the same, we again traverse backward the routine instance from the \emph{paste} UI until we find a \emph{copy} UI $u_3$ (line~\ref{alg:startPasteCheck} to~\ref{alg:stopPasteCheck}).\footnote{Our filtering approach, described in Section~\ref{sec:segmentation} guarantees that there exists a $u_3$ of type \emph{copy} preceding the \emph{paste} UI}. Then, we retrieve the target element of $u_3$ and we append it to queue $S$, and we add the copied value of $u_3$ to queue $I$ (lines~\ref{alg:addSource1} and \ref{alg:addInput1}).

For each encountered $u_2$ of type \emph{edit} (line~\ref{alg:checkForEdit}), we check its target element and we compare it to the target element of $u_1$. If they are the same (line~\ref{alg:sameEditTarget}), we push the \emph{target element} of $u_2$ to the front of queue $S$, and we push the \emph{data content} of the target element after the editing performed by $u_2$ to the front of the queue $I$ (line~\ref{alg:addSource2} and~\ref{alg:addInput2}). When we reach this point, we also stop the iteration over all the UI instances preceding $u_1$ because the value of the target element after performing $u_1$ is obtained from the last \emph{edit} UI performed on the same target element and any other UI (i.e., \emph{paste} UIs) between $u_2$ and $u_1$.

Finally, before moving to the next routine instance (i.e., returning to line~\ref{alg:traverseRoutineInstances}), we store the input data and the output data observed in the current routine instance for the normalized UI $\nui$ in the set $T$, which collects all the input and output data observed for \emph{all} the instances of $\nui$ (see line~\ref{alg:addTransformationExample}).

After performing all the above steps for each routine instance $r \in \riset$, and collecting all the required data to identify a possible data transformation function into the sets $T, K,$ and $\Pi$, we look for the data transformation function by leveraging two state-of-the-art tools:
Foofah~\cite{DBLP:conf/sigmod/JinACJ17} and TANE~\cite{DBLP:journals/cj/HuhtalaKPT99}. First, we try to identify the data transformation function using Foofah, then -- if Foofah fails -- we use TANE.

Foofah requires in input two series of data values, one referred to as \emph{input} and one referred to as \emph{output}. We generate the two series from the pairs $(I,O)$ that we collected in $T$, which capture examples of data transformations. From these examples, Foofah tries to synthesize an optimal data transformation function to convert input(s) to output.\footnote{For more details about Foofah refer to~\cite{DBLP:conf/sigmod/JinACJ17}.} We note that we run Foofah under the assumption that the output series is noise- and error-free, i.e., the analyzed data transformations are supposed to be correct.

However, Foofah suffers from two limitations: it is inefficient when the size of the input and output series is large; it cannot discover conditional data transformation functions (where different manipulations are applied depending on the input). Hence Foofah cannot deal with heterogeneous data.
To address these limitations, we group the data transformation examples into equivalence classes, where each class represents a different structural pattern of the input data. To create these equivalence classes, for each data sample in the input data series, we discover its symbolic representation describing its structural pattern by applying \emph{tokenization}. The tokenization that we apply replaces each maximal chained subsequence of symbols of the same type (either digits or letters) with a special token character ($\langle d \rangle+$ or $\langle a \rangle+$, resp.), and leaves any other symbol unaltered. For each equivalence class, we discover a data transformation function by providing to Foofah one randomly selected data transformation example from the equivalence class. The use of equivalence classes allows us to remove the heterogeneity of the input data and to facilitate the application of Foofah, which will operate only on a single data transformation example.



If Foofah cannot identify a data transformation function (line~\ref{alg:syntTransDiscoveryEnd}), we turn to TANE, which can discover semantical data transformation functions (also known as \emph{functional dependencies}~\cite{DBLP:journals/cj/HuhtalaKPT99}). TANE requires in input a table where each row contains $n-1$ input data values and an output data value in column $n$ (this is conceptually similar to the input and output series required by Foofah). TANE analyzes each row of such a table to check if there exists any dependency between the values in the first $n-1$ columns and the value in column $n$.\footnote{For more details about TANE refer to~\cite{DBLP:journals/cj/HuhtalaKPT99}.} An example of a semantical data transformation function discovered by TANE would be: if the value of column $i$ is X, then the value of column $n$ is always Y.

In our context, the input table for TANE is a table where each row represents the output data observed in all the UIs preceding $\nui$ in a routine instance, and the last element of the row is the output data of the $\nui$ instance in that routine (i.e., the value of the element edited by the execution of $\nui$ in that routine instance). To build such a table, we require in input all the instances of $\nui$ (which we stored in the set $K$) as well as all the instances of any UI preceding $\nui$ (which we stored in the set $\Pi$). If TANE identifies a semantical data transformation function (line~\ref{alg:depfound}), we set $\nui$ as deterministic (through the boolean $d$), and we compose the data transformation step using the output of TANE (see lines~\ref{alg:tane1} to~\ref{alg:tane2}).

Table~\ref{table:dependencyTable} shows an example of the dependency table that we would build from the log captured in Table~\ref{tab:uiLog} (assuming that the full-length UIs log contains nine instances of the routine showed in rows 1 to 24). Giving Table~\ref{table:dependencyTable} in input to TANE, it would identify that the value of the last column (i.e., the type of student, domestic or international) can be deterministically generated by observing the value of column four (i.e., \emph{country of residence}).

\begin{table}[tbh]
\centering
\scriptsize{
\begin{tabular}{c|c|c|c|c}
\hline
Full name & Date & Phone & Country of residence & Target \\ \hline
Albert Rauf & 11-04-1986 & 043-512-4834 & Germany & International \\
John Doe & 11-03-1986 & 024-706-5621 & Australia & Domestic \\ 
Steven Richards & 18-06-1986 & 088-266-0827 & Australia & Domestic \\ 
Hilda Diggle & 31-07-1993 & 073-672-5593 & New Zealand & International \\ 
Luca Bianchi & 19-10-1998 & 029-211-4904 & Italy & International \\ 
Igor  & 13-08-1993 & 040-656-3417 & Ukraine & International \\ 
Ben Stanley & 03-12-1991 & 244-557-2104 & Australia & Domestic \\ 
Olga Mykolenchuk & 11-04-2000 & 956-045-0703 & Ukraine & International \\ 
Daniel Brown & 06-04-1994 & 032-660-0403 & New Zealand & International \\ \hline
\end{tabular}
}
\caption{Example of a dependency table.}
\label{table:dependencyTable}
\end{table}

If also TANE does not discover any data transformation function, it means that we are not able to automatically determine the value of the element edited by the execution of $\nui$, consequently we assume that $\nui$ is not deterministic.  Otherwise, we output the data transformation step discovered.


\begin{figure}[htb]
\centering
\hspace*{-1cm}
\includegraphics[scale = 0.7]{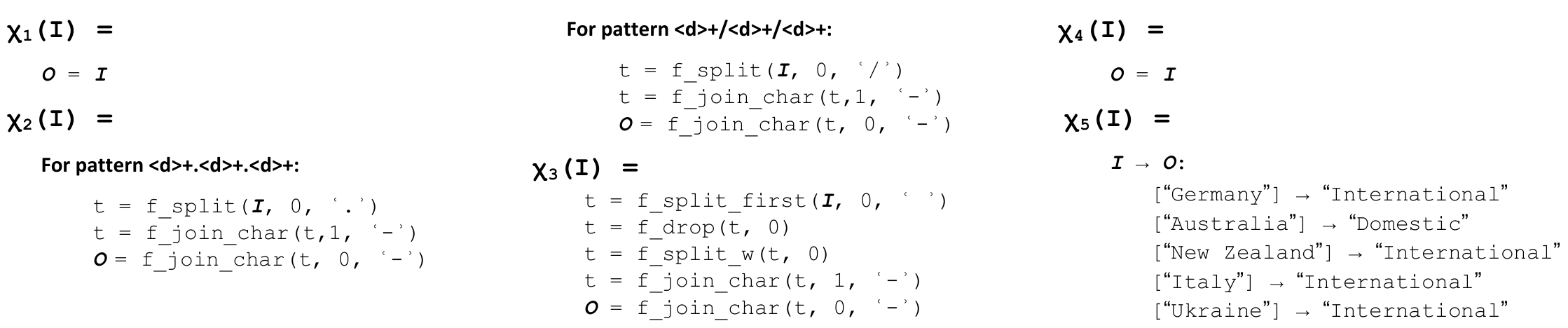}
\caption{Transformation functions discovered from the running example}
\label{fig:tfunctions}
\end{figure}


\begin{table}[tbh]
\makebox[0.98\textwidth][c]{
\centering
{\scriptsize{
\begin{tabular}{c|c|c|c}

\textbf{Transformation step} & \textbf{Sources} & \textbf{Target} & \textbf{Transformation function}\\\hline
$\lambda_1$ & Cell A& Full Name & $\chi_1$\\\hline
$\lambda_2$ &  Cell B & Date & $\chi_2$\\\hline
$\lambda_3$ &  Cell C & Phone & $\chi_3$\\\hline
$\lambda_4$ &  Cell D & Country & $\chi_4$\\\hline
$\lambda_5$ & Country & Status & $\chi_5$\\\hline
\end{tabular}
}}}
\caption{Transformation steps}
\label{tab:transSteps}
\end{table}

Considering our running example, Figure~\ref{fig:tfunctions} shows the data transformations functions discovered by Foofah (t1 to t4) and by TANE (t5) when running Algorithm~\ref{alg:checkeditUIs} on an hypothetical extended version of the UI log in Table~\ref{tab:uiLog} and giving as input the routine shown in rows 1 to 24 (Table~\ref{tab:uiLog}) along all its instances, and the \emph{edit} UIs at rows 6, 11, 16, 21, 23 (respectively, for identifying the data transformation functions from t1 to t5). Each data transformation function shows how input data is turned into output data. Although some rules are intuitive to interpret (e.g., t1 and t5), others may appear slightly cryptic. We refer to Foofah~\cite{DBLP:conf/sigmod/JinACJ17} and TANE~\cite{DBLP:journals/cj/HuhtalaKPT99} original studies for an extensive description of the set of rules that the two tools are capable to discover.

Finally, the data transformation functions are integrated into the data transformation steps, which also include the instantiation of the input and the output of the function, as shown in Table~\ref{tab:transSteps}. 

\subsection{Routines aggregation}
\label{sec:routinesAggregation}
When a routine can be performed by executing a set of UIs without following a strict order, we may observe multiple execution variants of the same routine in the log. For example, if a worker needs to copy the \emph{first name}, the \emph{family name}, and the \emph{phone number} of a set of customers from a spreadsheet to different web-forms, she may choose to copy the data of each customer in any order (e.g., \emph{first name}, \emph{phone number}, and \emph{family name}, or \emph{family name}, \emph{phone number}, \emph{first name}). In such a scenario, the UI log would record several different execution variants of the same routine. Routine execution variants do not bring any additional value, rather they just generate redundancy within the log leading to the discovery of different routine specifications that would actually execute (once deployed as software bots) the same routine. Considering these routine specification as duplicates, this final step focuses on their removal.

\begin{figure}[htb]
\centering
\hspace*{-1cm}
\includegraphics[scale = 0.9]{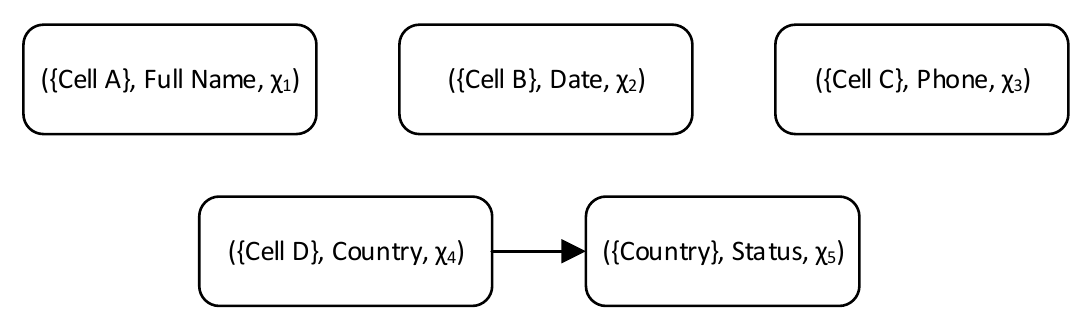}
\caption{Data transformation graph example}
\label{fig:transformationGraph}
\end{figure}

To identify duplicate routine specifications, we start by generating for each routine discovered in the previous step its \emph{data transformation graph}.
\begin{definition}[\textbf{Data Transformation Graph}]
Given a routine specification ($c_i$, $\Lambda$), its \emph{data transformation graph} is a graph $G_\Lambda = (D_\Lambda , L_\Lambda)$, where:
$D_\Lambda $ is the set of vertices of the graph, and each vertex $d \in D_\Lambda$ maps one data transformation step $\lambda \in \Lambda$;
$L_\Lambda  \subseteq D_\Lambda  \times D_\Lambda$ is the set of edges of the graph, and each edge $(d_i, d_j) \in L_\Lambda$ represents a dependency between two data transformation steps capturing the fact that the target of the data transformation step mapped by $d_i$ is (one of) the source(s) of the data transformation step mapped by $d_j$.
\end{definition}
\figurename~\ref{fig:transformationGraph} shows the data transformation graph of the routine we discovered in the previous step in our running example.

Data transformation graphs can be used to check whether two routine specifications are equivalent, in fact, two routine specifications, ($c_i$, $\Lambda_1$) and ($c_j$, $\Lambda_2$), are equivalent if and only if the following two relations hold: i) their data transformation graphs are the same, i.e., $D_{\Lambda_1}$ = $D_{\Lambda_2}$ and $L_{\Lambda_1}$ = $L_{\Lambda_2}$; ii) their candidate routines $c_i$ and $c_j$ contain the same set of UIs, and all the UIs of type \emph{click button} appear in the same order in both $c_i$ and $c_j$.



By comparing each pair of routine specifications, we first create sets of equivalent routine specifications, and, for each set, we discard all the routine specifications but one. Ideally, we would like to retain the best routine specification of each set, however, we need to define what it means to be the \emph{best} one. We can select the best routine specification by relying on different quantitative metrics, such as frequency, length, or duration of the candidate routine of a routine specification. For example, we can choose frequency as a selection criterion and retain from each set the routine specification whose candidate routine is the most frequent in the UI log. 

Intuitively, the most frequent candidate routine represents the common routine execution, so that one may be tempted to use that criterion by default. However, the most frequent routine execution is not necessarily the optimal execution. For example, length or duration could represent better selection criteria. Length prioritizes short candidate routines over long ones, assuming that a candidate routine should comprise as few steps as possible. Duration prioritizes execution times over the number of steps. The duration of a candidate routine can be estimated as the average execution time of each routine instance of the candidate routine that is recorded in the UI log. Note, however, that the duration could be not always reliable since during the routine execution, the worker might perform activities that do not appear in the log or that are not relevant for the routine execution, thus involuntarily increasing the observed execution time of the routine. For this reason, we implemented a combination of length and frequency to select the best routine specification from each set. Precisely, we use length first and then compare the frequencies of the candidate routines having the same length.

\section{Evaluation}
\label{sec:evaluation}

\urldef{\footurla}\url{https://github.com/volodymyrLeno/RPM_Miner}
\urldef{\footurlb}\url{https://doi.org/10.6084/m9.figshare.12543587}

We implemented our approach as an open-source Java command-line application\footnote{Available at \footurla} and also embedded this in the open-source tool Robidium~\cite{LenoDPRDM20}. Using the command-line application, we conducted a series of experiments to analyze the applicability of our approach in real-life settings.
Specifically, we assessed to what extent our approach can rediscover routines that are known to be recorded in the input UI logs,
and analyzed whether our approach is able to correctly identify automatable and not automatable user interactions within such routines.

Accordingly, we define the following research questions:
\begin{itemize}
\item \textbf{RQ1.} Does the approach discover candidate routines that are known to exist in a UI log?
\item \textbf{RQ2.} Does the approach discover automatable routines that are known to be present in a UI log?

\end{itemize}

\subsection{Datasets}
\label{sec:datasets}


To answer our research questions, we rely on a dataset of 13 logs. These logs can be divided into three subgroups: artificial logs, real-life logs recorded in a supervised environment, and real-life logs recorded in an unsupervised environment.\footnote{The real-life logs were recorded with the Action Logger tool~\cite{DBLP:conf/bpm/LenoPRDM19}. All the logs are available at \footurlb}
Table~\ref{table:datasets} shows the logs characteristics.

\begin{table}[tbh]
\centering
\scalebox{0.85}{
\begin{tabular}{l|c|c|c|c}
\hline
\textbf{UI Log} & \textbf{\# Routine} & \textbf{\# Task}  & \textbf{\# Actions} & \textbf{\# Actions per} \\ 
& \textbf{Variants} & \textbf{Traces}  &  & \textbf{trace (Avg.)} \\ \hline
CPN1 & 1 & 100 & 1400 & 14.000 \\ 
CPN2 & 3 & 1000 & 14804 & 14.804 \\ 
CPN3 & 7 & 1000 & 14583 & 14.583 \\ 
CPN4 & 4 & 100 & 1400 & 14.000 \\ 
CPN5 & 36 & 1000 & 8775 & 8.775 \\ 
CPN6 & 2 & 1000 & 9998 & 9.998 \\ 
CPN7 & 14 & 1500 & 14950 & 9.967 \\ 
CPN8 & 15 & 1500 & 17582 & 11.721 \\ 
CPN9 & 38 & 2000 & 28358 & 14.179 \\ 
Student Records (\studentRecord) & 2 & 50 & 1539 & 30.780 \\
Reimbursement (\reimbursement) & 1 & 50 & 3114 & 62.280 \\ 
Scholarships 1  (\scholarshipA) & & - & 693 &  \\ 
Scholarships 2  (\scholarshipB) & & - & 509 & \\ \hline
\end{tabular}
}
\caption{UI logs characteristics.}
\label{table:datasets}
\end{table}

The artificial logs (CPN1--CPN9) were generated from Colored Petri Nets (CPNs) in \cite{bosco2019}.
The CPNs used have increasing complexity, from low (the net used to generate CPN1) to high (the net used for CPN9).
The underlying routines are characterized by a varying amount of non-deterministic user interactions injected.
They involve simple data transformations, mostly in the form of copy-pasting.
The logs generated were originally noise-free and segmented. We removed the segment identifiers to produce unsegmented logs.

The \emph{Student Records} (\studentRecord) and \emph{Reimbursement} (\reimbursement) logs record the simulation of real-life scenarios.
The \studentRecord \ log simulates the task of transferring students' data from a spreadsheet to a Web form.
The \reimbursement \ log simulates the task of filling reimbursement requests with data provided by a claimant.
Each log contains fifty recordings of the corresponding task executed by one of the authors, who followed strict guidelines on how to perform the task.
These logs contain little noise, which only accounts for user mistakes,
such as filling the form with an incorrect value and performing additional actions to fix the mistake.
For both logs, we know how the underlying task was executed, and we treat such information as ground truth when evaluating our approach.
While the routines captured in the logs are fully automatable, they include complex transformations to test the automatability assessment step of the approach.

Finally, the \emph{Scholarships} logs (\scholarshipA \ and \scholarshipB) were recorded by two employees of the University of Melbourne who performed the same task. It is the task of processing scholarship applications for international and domestic students.
This task mainly consists of students' data manipulation with transfers between spreadsheets and Web pages.
Compared to the other logs used in our experiences, we have no a-priori knowledge of how to perform the task at hand (no ground truth).
Also, when recording the logs, the University employees were not instructed to perform their task in a specific manner,
i.e.,\ they were left free to perform this task as they would normally do when unrecorded.

\subsection{Setup}
\label{sec:setup}

To measure the quality of the discovered candidate routines, we use the Jaccard Coefficient (JC),
which captures the level of similarity between discovered and ground truth routines.
JC does not penalize the order of the interactions in a routine, which follows from the assumption that a routine could be executed by performing some actions in a different order.
The JC between two routines is the ratio $\frac{n}{m}$,
where $n$ is the number of user interactions that are contained in both routines,
while $m$ is the total number of user interactions present in the two routines.

Given the set of discovered routines and the set of ground truth routines,
for each discovered routine, we compute its JC with all the ground truth routines and assign the maximum JC to the discovered routine as its quality score.
Finally, we assess the overall quality of the discovered routines as the average of the JC of each discovered routine. As the ground truth, we use the segments of the artificial logs and the guidelines given to the author who performed the tasks in \studentRecord \ and \reimbursement.

The JC alone is not enough to assess the quality of the discovered routines,
as this measure does not consider the routines we may have missed in the discovery.
Thus, we also measure the total coverage to quantify how much log behavior is captured by the discovered routines.
We would like to reach high coverage with as few routines as possible.
Thus, we prioritize long routines over short ones by measuring the average routine length alongside its coverage.

We assess the quality of the automatable routines discovery by measuring precision, recall and F-score.
For each discovered routine, we compute the corresponding confusion matrix, where \emph{true positives} (TP) are correctly identified automatable user interactions,
\emph{true negatives} (TN) are correctly identified non-automatable user interactions, \emph{false positives} (FP) are the user interactions that were wrongly marked as automatable,
and \emph{false negatives} (FN) are the user interactions that were wrongly marked as non-automatable.
From the constructed confusion matrix, we calculate precision, recall and F-score as follows:

\begin{equation}Precision = \frac{TP}{TP + FP},\end{equation}
\begin{equation}Recall = \frac{TP}{TP + FN},\end{equation}
\begin{equation}F{\text -}score = 2 \cdot \frac{Precision \cdot Recall}{Precision + Recall}.\end{equation}

We report the averages of these metrics for all the discovered routines in the log.
We also report the average ratio of automatable user interactions for the routines in the log.

The results for the \scholarshipA \ and \scholarshipB \ logs were qualitatively assessed with the help of the University of Melbourne employees who performed the task. Specifically, we asked them to compare the rediscovered routines with the actions they performed while recording.

All experiments were conducted on a Windows 10 laptop with an Intel Core i5-5200U CPU 2.20 GHz and 16GB RAM,
using cohesion as a routine selection criterion with the minimum support threshold set to 0.1 and the minimum coverage threshold equal to 0.05.

\subsection{Results}

Table~\ref{table:candidatesIdentification} shows the quality of the discovered routine candidates. Although the synthetic logs only contain the user interactions that belong to routines, we achieved perfect coverage for three logs only, namely CPN1, CPN4 and CPN6.
This is because some execution patterns were observed very rarely.
Since the \studentRecord \ and \reimbursement \ logs contain noise, the coverage cannot be 1 in these two cases.
For six logs out of eleven logs, the discovered routines match with the ground truth.
Overall, the JC is very high, above 0.95 for all the logs except CPN5.
The underlying model of the CPN5 log consists of multiple branches, generating 36 different executions.
Considering the fact that some execution patterns are not frequent enough, we discovered only partial routines.
As can be seen clearly, for this log we also achieved the lowest coverage (0.84).
For the \reimbursement \ log we found two routines consisting of an identical set of actions.
These routines were not merged though, because they are characterized by different transformation functions.


\begin{table}[tbh]
\centering
\scalebox{0.85}{
\begin{tabular}{l|c|c|c|c|c}
\hline
\textbf{UI Log} & \textbf{\# Routines} & \textbf{Length}  & \textbf{ Length} & \textbf{Total} & \textbf{JC} \\ 
& \textbf{discovered} & \textbf{(Max)}  & \textbf{(Avg.)} & \textbf{coverage} & \\ \hline
CPN1 & 1 & 14 & 14.00 & 1.00 & 1.000 \\ 
CPN2 & 2 & 15 & 14.50 & 0.95 & 1.000 \\ 
CPN3 & 3 & 19 & 14.33 & 0.93 & 1.000 \\ 
CPN4 & 4 & 14 & 14.00 & 1.00 & 1.000 \\ 
CPN5 & 8 & 8 & 7.38 & 0.84 & 0.880 \\ 
CPN6 & 2 & 11 & 10.00 & 1.00 & 1.000 \\ 
CPN7 & 7 & 10 & 9.43 & 0.93 & 0.971\\  
CPN8 & 6 & 18 & 10.67 & 0.91 & 0.967\\ 
CPN9 & 6 & 18 & 14.67 & 0.95 & 1.000 \\ 
\studentRecord & 2 & 31 & 30.00 & 0.917 & 0.967 \\
\reimbursement & 2 & 61 & 61.00 & 0.903 & 0.967 \\ \hline
\end{tabular}
}
\caption{Candidates identification}
\label{table:candidatesIdentification}
\end{table}

Table~\ref{table:automatableResults} shows the quality of the automatable routines discovery.
We correctly identified all the automatable and not automatable user interactions for the CPN3, CPN6 and \studentRecord \ logs.
The routines recorded in the CPN3 and \studentRecord \ logs are fully automatable.
Although the \reimbursement \ log contains automatable routines only, our approach failed to discover some of the underlying transformations,
and, therefore, incorrectly marked some interactions as not automatable.
Some of the user interactions of the synthetic logs were wrongly identified as automatable.
Although the data values of such interactions can be deterministically computed, the locations of the edited elements were completely random as it was intended in the corresponding models.
Thus, in practice, such interactions are not automatable.
The routines discovered from the CPN5 log are characterized by the lowest number of automatable user interactions, and we achieved the lowest recall for this log (0.805).
Overall, F-score is high, above 0.85 for all the logs, except CPN7 and CPN8.
For these logs we also achieved the lowest recall, meaning that some interactions of the corresponding routines were wrongly identified as not automatable.
Although in the CPN models used to generate the artificial logs, some of the interactions are not deterministic, they are automatable in the context of the discovered routines.
For example, for the CPN9 log we discovered six routines that correspond to the different branches within the model.
For all the executions of a branch we use the same data values, and hence, the corresponding user interactions are automatable.


\begin{table}[tbh]
\centering
\scalebox{0.85}{
\begin{tabular}{l|c|c|c|c}
\hline
\textbf{UI Log} & \textbf{RAI} & \textbf{Precision} & \textbf{Recall}  & \textbf{F-score} \\ 
& \textbf{(Avg.)} & \textbf{(Avg.)} & \textbf{(Avg.)} &  \textbf{(Avg.)} \\ \hline
CPN1 & 1.000 & 0.928 & 1.000 & 0.963 \\ 
CPN2 & 0.931 & 0.926 & 1.000 & 0.961 \\ 
CPN3 & 1.000 & 1.000 & 1.000 & 1.000 \\ 
CPN4 & 0.786 & 1.000 & 0.846 & 0.917 \\ 
CPN5 & 0.728 & 0.812 & 1.000 & 0.896 \\ 
CPN6 & 0.742 & 1.000 & 1.000 & 1.000 \\ 
CPN7 & 0.546 & 0.907 & 0.805 & 0.841 \\ 
CPN8 & 0.612 & 0.897 & 0.823 & 0.845 \\ 
CPN9 & 0.741 & 0.951 & 0.886 & 0.916 \\ 
\studentRecord & 1.000 & 1.000 & 1.000 & 1.000 \\
\reimbursement & 0.967 & 1.000 & 0.967 & 0.983 \\ \hline
\end{tabular}
}
\caption{Automatable routines discovery}
\label{table:automatableResults}
\end{table}


From the \scholarshipA\ log we discovered five fully automatable routines.
The first routine consists in manually adding graduate research student applications to the student record in the university's student management system.
The application is then assessed, and the student is notified of the outcome.
The second routine consists in lodging a ticket to verify possible duplicate applications.
When a new application is entered in the system and its data matches an existing application,
the new application is temporarily put on hold, and the employee fills in and lodges a ticket to investigate the duplicate.
The remaining three routines represent exceptional cases, where the employee either executed the first or the second routine in a different manner (i.e.,\ by altering the order of the actions or overlapping routines executions).
These routines were not identified as duplicate because they are characterized by different sequences of button clicks.

To assess the results, we showed the discovered routines to the employee of the University of Melbourne
who recorded the \scholarshipA\ log, and they confirmed that the discovered routines correctly capture their task executions.
Also, they confirmed that the last three routines are alternative executions of the first routine.\footnote{Detailed results at \footurlb}

While the results from the \scholarshipA \ log were positive,
our approach could not discover any correct routine from the \scholarshipB \ log.
By analyzing the results, we found out that the employee worked with multiple worksheets at the same time,
frequently switching between them for visualization purposes.
Such behavior recorded in the log negatively affects the construction of the CFG and its domination tree,
ultimately leading to the discovery of incorrect segments and routines.


Table~\ref{table:executionTimes} shows the execution time for each step of the approach.
As we can see, the most computationally heavy step is the automatability assessment.
For all the logs, this step took the largest amount of time, except for the CPN5, \scholarshipA, and \scholarshipB \ logs.
While the execution time is still reasonably low for all the artificial logs, it substantially increases for the \studentRecord \ and \reimbursement \ logs.
In these two logs, the automatability assessment took 99 percent of the total computation time.
This is caused by the fact that the underlying transformations in these two logs were very complex, often involving regular expressions or long sequences of manipulations.
In contrast, all the transformations in the CPN1-CPN9 logs were simple copy-paste operations.
Overall, for the synthetic logs, the approach took no more than 42 seconds.
The aggregation step required the smallest amount of time.
For the CPN1 log, we discovered only one routine, and, therefore, we did not have to apply any aggregation.
For the \scholarshipA \ and \scholarshipB logs, the most time taking step was the segmentation. 
The CFGs constructed for these logs were very complex, with a high number of loops. 
This significantly increased the time to identify back-edges in such CFGs and, therefore, the total time of segmentation. 

\begin{table}[tbh]
\centering
\scalebox{0.85}{
\begin{tabular}{l|c|c|c|c|c}
\hline
& \multicolumn{5}{|c}{\textbf{Execution time (sec)}}
\\\cline{2-6}

\textbf{UI Log}
& \textbf{Segmentation}
& \textbf{Candidates}
& \textbf{Automatability}
& \textbf{Aggregation}
& \textbf{Total}\\

& & \textbf{identification} & \textbf{assessment} & & \\\hline

CPN1 & 0.337 & 2.766 & 7.148 & -- & 10.251 \\ 
CPN2 & 2.521 & 4.482 & 17.408 & 0.010 & 24.421 \\ 
CPN3 & 1.570 & 9.781 & 15.545 & 0.010 & 26.906 \\ 
CPN4 & 0.637 & 5.013 & 18.409 & 0.009 & 24.068 \\ 
CPN5 & 1.169 & 20.223 & 19.761 & 0.008 & 41.161 \\ 
CPN6 & 1.376 & 3.811 & 6.102 & 0.010 & 11.299 \\ 
CPN7 & 2.497 & 15.196 & 17.594 & 0.010 & 35.297 \\ 
CPN8 & 2.469 & 14.605 & 17.399 & 0.010 & 34.483 \\ 
CPN9 & 2.877 & 12.272 & 18.798 & 0.013 & 33.960 \\ 
\studentRecord & 0.801 & 8.077 & 845.255 & 0.013 & 854.146 \\ 
\reimbursement & 2.022 & 8.657 & 1066.041 & 0.011 & 1076.731 \\ 
\scholarshipA & 29.052 & 14.066 & 21.400 & 0.011 & 64.529 \\ 
\scholarshipB & 403.903 & 152.474 & -- & -- & 556.377 \\\hline
\end{tabular}
}
\caption{Execution time}
\label{table:executionTimes}
\end{table}

\subsection{Threats to validity}

The reported evaluation has a number of threats to validity. First, a potential threat to internal validity is the fact that the context parameters (i.e.\ the attributes in the log that capture the notion of ``user interaction'') were manually selected. These context parameters are required as one of the inputs of the proposed method (in addition to the UI log). To mitigate this threat, the parameters were first selected by each of the two authors of the paper independently, then cross-checked to reach a mutual agreement, and then validated by the other authors based on their understanding of the event logs in question. 
 
Another possible threat to internal validity is the limited use of parameter values to configure the approach at hand. To ensure we do not miss any significantly important behavior in the logs, we used very low support and coverage, equal to 0.1 and 0.05, respectively. 

A potential threat to external validity is given by the use of a limited number of real-life logs (four). 
These logs focus on one type of task that can be automated via RPA, namely data transferring. 
These logs, however, exhibit different characteristics in terms of the complexity of the captured processes and log size. 
To mitigate this threat, we additionally performed a more extensive evaluation on a battery of artificial logs. 
For two real-life logs, we had no information about the underlying processes.
Therefore we evaluated the results qualitatively with the workers responsible for their execution.
To ensure the full reproducibility of the results, we have released all the logs, both real-life and artificial, used in our experiments.
The only exceptions are the \scholarshipA \ and \scholarshipB \ logs as they contain sensitive information.

\section{Conclusion}
\label{sec:conclusion}
\medskip

This paper presented an approach to discover automatable routines from UI logs. 
The approach starts by decomposing the UI log into segments corresponding to paths within the connected components of a control-flow graph derived from the log. These paths represent sequences of actions that are repeated multiple times within the event log, possibly with some variations. 
Once the log is segmented, a noise-resilient sequential pattern mining technique is used to extract frequent patterns that corresponds to the candidate routines.
Next, the candidate routines are assessed for their amenability to automation. For each routine, a corresponding executable specification is synthesized, which can be compiled into an RPA script.
Finally, the approach identifies semantically equivalent routines in order to produce a non-redundant set of automatable routines.
 
The approach has been implemented as an open-source tool, namely Robidium. This article reported on an evaluation of the fit-for-purpose and computational efficiency of the proposed approach. The evaluation shows that the approach can rediscover routines injected into synthetic logs, and that it discovers relevant routines in real-life logs. For most logs, the execution time does not exceed one minute. The only exceptions related to logs where we deliberately injected complex data transformations or where the routine instances overlap in the UI log. 


The proposed approach makes a number of limiting assumptions.
First, the effectiveness of the approach is sensitive to noise, e.g.\ clicks that are not related to the routine itself or clicks resulting from user mistakes. 
In our evaluation, we observed this phenomenon to varying degrees when dealing with real-life logs.
In practice, the approach can identify correct routines only if they are frequently observed in the log.
Recurring noise affects the accuracy of the results. To address this limitation, we will investigate the use of alternative segmentation and sequential pattern discovery techniques that incorporate noise tolerance mechanisms. Another avenue is to discover sequential patterns using the approach outlined in this article and then to filter out patterns that are \emph{chaotic} in the sense that their occurrence does not affect the probability of other patterns occurring subsequently nor vice-versa. This latter approach has been studied in the context of event log filtering for process mining in~\cite{DBLP:journals/jiis/TaxSA19}.

Second, the approach is designed for logs that capture consecutive routine executions. In practice, routine instances may sometimes overlap (cf. the \scholarshipB \ real-life log in the evaluation). A possible avenue to address this limitation is to search for overlapping frequent patterns directly in the unsegmented log, instead of first segmenting it and then finding patterns in the segmented log. This approach has been previously investigated in the context of so-called Local Process Mining (LPM), where the goal is to discover process models capturing frequently repeated (and possibly overlapping) behavior in an unsegmented sequence of events~\cite{LPM}. 

When assessing the automatability of a routine, the propsoed approach assumes that the values of the edited fields are entirely derived
from the (input) fields that are explicitly accessed (e.g., via copy operations) during the routine's execution.
Hence, it will fail to identify automatable user interactions in the case where a worker visually reads from a field
(without performing a \emph{copy} operation on it) and writes what they see into another field. An avenue for addressing this  limitation is to complement the proposed method with optical character recognition techniques over screenshots taken during the UI log recording, so as to be able to detect that some of the outputs of a routine come from fields that have not been explicitly accessed via a copy-to-clipboard operation.

Furthermore, the proposed approach is unable to discover conditional behavior, where the transformation function for the target field depends on the value of another field. Consider, for example, a routine that involves copying delivery data.
If the delivery country is USA, then the month comes before the day (MM/DD/YYYY), otherwise the day comes before the month.
Here, the transformation function depends on a condition of the form  ``country = USA'', which the proposed approach is unable to discover.
In a similar vein, the proposed approach is able to discover transformations that depend on the structural pattern of the value of the input field(s), but it fails to distinguish the patterns that, although having the same syntactical structure, have different semantics. Following the example above, our approach will put both date types into the same equivalence class. Addressing this limitation would require the development of more sophisticated data transformation discovery techniques, beyond the capabilities of Foofah.


Finally, the method to detect if two routines are semantically equivalent assumes that all button clicks in a UI are effectful, meaning that their presence and the order in which they occur affect the outcome of the routine. 
In practice, some clicks may have no effect on the routine's outcome. For example, some clicks may simply serve to pop up a help box, while others may just serve to move from one page to another in a listing. 
To address this limitation, we foresee extensions of the proposed method where the alphabet of the UI log is extended with a richer array of actions, and where the routine discovery approach can be configured via a language for the specification of action effects. 



\medskip\noindent\textbf{Acknowledgments}. The authors thank Stanislav Deviatykh for his help in the prototype implementation. This research is supported by the Australian Research Council (DP180102839) and the European Research Council (project PIX).

\vspace{-2mm}
\bibliographystyle{elsarticle-num}
\bibliography{lit}

\end{document}